\documentclass{aa}
\usepackage{graphicx}
\usepackage{txfonts}
\usepackage{multirow}
\usepackage[colorlinks=true, linkcolor=blue, urlcolor=blue, citecolor=blue]{hyperref}

\begin{document}

\title{Orbit-based structural decomposition and stellar population recovery for edge-on barred galaxies}
\titlerunning{Structural decomposition and stellar population recovery for edge-on barred galaxies}

\author
{Yunpeng Jin\inst{1}\thanks{E-mail: jyp199333@163.com}
\and Ling Zhu\inst{2}\thanks{E-mail: lzhu@shao.ac.cn}
\and Behzad Tahmasebzadeh\inst{3,4}
\and Shude Mao\inst{1}
\and Glenn van de Ven\inst{5}
\and Timothy A. Davis\inst{6}}

\institute
{Department of Astronomy, Westlake University, Hangzhou, Zhejiang 310030, China
\and Shanghai Astronomical Observatory, Chinese Academy of Sciences, 80 Nandan Road, Shanghai 200030, China
\and Department of Astrophysics and Planetary Science, Villanova University, 800 E Lancaster Ave, Villanova, PA 19085, USA
\and Department of Astronomy, University of Michigan, Ann Arbor, MI, 48109, USA
\and Department of Astrophysics, University of Vienna, Türkenschanzstraße 17, 1180 Wien, Austria
\and Cardiff Hub for Astrophysics Research \& Technology, School of Physics \& Astronomy, Cardiff University, Queens Buildings, The Parade, Cardiff, CF24 3AA, UK}
\date{}

\abstract
{In our previous paper, we developed an orbit-superposition method for edge-on barred galaxies and constructed a set of dynamical models based on different mock observations of three galaxies from the Auriga simulations. In this study, we adopted 12 cases with side-on bars (three simulated galaxies, each with four different projections). We decomposed these galaxies into different structures combining the kinematic and morphological properties of stellar orbits. We then compared the model-predicted components to their true counterparts in the simulations. Our models can identify (BP/X-shaped) bars, spheroidal bulges, thin discs, and spatially diffuse stellar halos. The mass fractions of bars and discs are well constrained with absolute biases of $|f_{\rm model}-f_{\rm true}|\le0.15$. We recovered the mass fractions of halos with $|f_{\rm model}-f_{\rm true}|\le0.03$. For the bulge components, 10 out of 12 cases exhibit $|f_{\rm model}-f_{\rm true}|\le0.05$, while the other two cases exhibit $|f_{\rm model}-f_{\rm true}|\le0.10$. Then, by tagging the stellar orbits with ages and metallicities, we derived the chemical properties of each structure. For the stellar ages, our models recovered the negative gradients in the bars and discs, but exhibited relatively larger uncertainties for age gradients in the bulges and halos. The mean stellar ages of all components were constrained with absolute biases $|t_{\rm model}-t_{\rm true}|\rm\lesssim1\,Gyr$. For stellar metallicities, our models reproduced the steep negative gradients of the bars and bulges, as well as all different kinds of metallicity gradients in the discs and halos. Apart from the bulge in the simulated galaxy Au-18, the mean stellar metallicities of all other components were constrained with absolute biases of $|Z_{\rm model}-Z_{\rm true}|\rm\le0.5\,Z_{\odot}$.}


\maketitle
\begin{nolinenumbers}

\section{Introduction}
Our preliminary understanding of galaxy structures comes from their images. A typical spiral galaxy consists of a disc, a bar, and/or a bulge. Since the 1970s, different single-band surveys from ultraviolet (UV) to far-infrared (FIR) wavelengths have been carried out, providing information about the colour and star formation of these structures. These surveys have indicated that in spiral galaxies, bulges are typically red, gas-poor, and lack ongoing star formation, while discs are usually blue, gas-rich, and contain different star-forming regions \citep{Kennicutt1998}. Furthermore, the presence of a bar may enhance star formation in the central region of a galaxy (e.g. \citealp{Ho1997a,Ho1997b,Sakamoto1999}). These findings demonstrated the importance of analysing distinct structures independently to better understand galaxy formation and evolution.

In the past decades, significant progress has been made in decomposing galaxy structures. The most common approach, photometric decomposition, relies on the assumption that different components differ in their light distributions. Photometric decomposition was initially performed by fitting 1D surface brightness profiles, considering only two structural components: the bulge and the disc. The bulge component was modelled using the $R^{1/4}$ law \citep{deVaucouleurs1948} or a more generalised S\'ersic profile, $R^{1/n}$, with $n\gtrsim2$ \citep{Sersic1968}, while the disc component was characterised by an exponential profile (e.g. \citealp{Freeman1970,Kormendy1977,Burstein1979,Boroson1981,Kent1985}). Since the 1990s, high-resolution telescopes such as the Hubble Space Telescope and large-scale sky surveys such as the Sloan Digital Sky Survey (SDSS; \citealp{York2000}) have provided galaxy images across various wavelengths. These images greatly advanced the development and application of 2D photometric decomposition techniques (e.g. \citealp{Andredakis1995,deJong1996,Simard1998,Simard2002,Ratnatunga1999,Peng2002,Peng2010,deSouza2004,Laurikainen2004,Laurikainen2005,Gadotti2008,MendezAbreu2008,MendezAbreu2017,Kruk2018}), where the central bulge and bar are either considered as a single component or modelled separately. However, the results of photometric decomposition significantly depend on the assumed analytical functions of the light distributions, which might diverge from the truth. For example, discs are usually modelled with exponential profiles, whereas recent studies have demonstrated that discs can be down-bending and/or up-bending in their inner or outer regions \citep{MendezAbreu2017,Breda2020,Ding2023}; bars are often assumed to be prolate Ferrers bars (see \citealp{Binney2008}) or are characterised by S\'ersic profiles, while real bars can be triaxial with boxy, peanut, or X-shaped (hereafter BP/X-shaped) structures (e.g. \citealp{Shaw1987,Luetticke2000,Erwin2017,Li2017}). Furthermore, photometric decomposition provides limited information on the stellar populations of different components.

With the development of integral field spectroscopy (IFS) surveys in recent years, other decomposition techniques have become possible. Such IFS surveys as SAURON \citep{Bacon2001}, $\rm ATLAS^{3D}$ \citep{Cappellari2011}, CALIFA \citep{Sanchez2012}, SAMI \citep{Bryant2015}, MaNGA \citep{Bundy2015}, and multiple VLT/MUSE programmes \citep{Bacon2017} have provided spatially resolved spectra for thousands of nearby galaxies. Therefore, one option is to decompose galaxies directly by fitting their spectra (e.g. \citealp{Johnston2012,Johnston2014,Tabor2017,Tabor2019,Oh2020,Pak2021}) based on the full spectral fitting technique such as pPXF \citep{Cappellari2004,Cappellari2017}. This spectroscopic decomposition method is not only capable of separating different structures, but also estimating the stellar population of each component. Another approach is to decompose galaxies by combining the photometry and kinematics derived from the spectra (e.g. \citealp{Rigamonti2023,Rigamonti2024}). However, bar structures are not treated as individual components in these works.

Dynamical decomposition is an alternative technique. This technique is widely used in simulations as the 6D phase-space information of stellar particles is known (e.g. \citealp{Correa2017,Pillepich2019,RodriguezGomez2019,Du2019,Du2020,Pulsoni2020}). For real galaxies, dynamical decomposition can be achieved through the Schwarzschild's orbit-superposition method \citep{Schwarzschild1979,Schwarzschild1982,Schwarzschild1993}, which builds 3D galaxy models by fitting the luminosity distributions and stellar kinematics derived from IFS data. Several implementations of this method are currently in use (e.g. \citealp{vdB2008,Long2018,Vasiliev2020,Neureiter2021,Quenneville2022,Dattathri2024}). The \citet{vdB2008} implementation and its publicly available version DYNAMITE \citep{Jethwa2020}\footnote{\url{https://dynamics.univie.ac.at/dynamite_docs/}} were validated with simulated galaxies (e.g. \citealp{Zhu2018c,Jin2019}) and were used to analyse orbital structures for real galaxies in various IFS surveys (e.g. CALIFA, \citealp{Zhu2018a,Zhu2018b,Zhu2018c}; MaNGA, \citealp{Jin2020}; SAMI, \citealp{Santucci2022,Santucci2023}; $\rm ATLAS^{3D}$, \citealp{Thater2023}). This implementation was further updated to a population-orbit superposition method, which supports the modelling of stellar populations by fitting the ages and metallicities derived from IFS spectra (e.g. \citealp{Poci2019,Poci2021,Zhu2020,Zhu2022a,Zhu2022b,Ding2023,Jin2024}).

Based on the \citet{vdB2008} implementation, a barred orbit-superposition method was developed (\citealp{Tahmasebzadeh2021,Tahmasebzadeh2022}; see also the public code DYNAMITE) and applied to a barred galaxy NGC 4371, observed at an inclination angle $\theta\sim60^\circ$ \citep{Tahmasebzadeh2024}. This method is limited to non-edge-on cases ($\theta\lesssim80^\circ$) and stellar populations are not considered. The recently launched VLT/MUSE programme, GECKOS \citep{vdSande2024}, is designed to observe 36 Milky Way-like edge-on ($\theta\gtrsim85^\circ$) galaxies. It will provide high spatial resolution ($\rm<0.2\,kpc$) IFS data of these galaxies, including both stellar kinematics and stellar populations. Eight of the 12 galaxies in the first GECKOS internal data release are barred and have BP/X-shaped structures \citep{FraserMcKelvie2025}. Edge-on observations preserve more kinematic details of bar and disc rotations than non-edge-on cases, making them more effective for dynamical decomposition. In our previous paper \citep{Jin2025}, we extended the method developed by \citet{Tahmasebzadeh2021,Tahmasebzadeh2022} to model edge-on ($\theta\gtrsim80^\circ$) barred galaxies that may exhibit BP/X-shaped structures. Using three simulated galaxies from the Auriga simulations \citep{Grand2017,Grand2024} and creating a total of 16 mock datasets, the effectiveness of this method in estimating the bar pattern speed, $\rm\Omega_p$, was validated, achieving an accuracy of $\sim13\%$. In this paper, we perform orbit-based structural decomposition for edge-on barred galaxies based on the model results from \citet{Jin2025}. Then, by tagging the stellar orbits with ages and metallicities, we evaluate how accurately ages and metallicities can be recovered for different components.

The paper is organised as follows. In Sect.~\ref{sec2}, we introduce the simulated galaxies and subsequent mock data. In Sect.~\ref{sec3}, we demonstrate how we create barred population-orbit superposition models, including the orbit-based decomposition of galaxy structures. In Sects.~\ref{sec4} and~\ref{sec5}, we characterise the properties of model-decomposed structures and compare them with their true counterparts in simulations. We discuss our results in Sect.~\ref{sec6} and summarise our conclusions in Sect.~\ref{sec7}.

\section{Auriga simulated galaxies and mock data}
\label{sec2}
\begin{figure*}
    \centering
    \includegraphics[width=12cm]{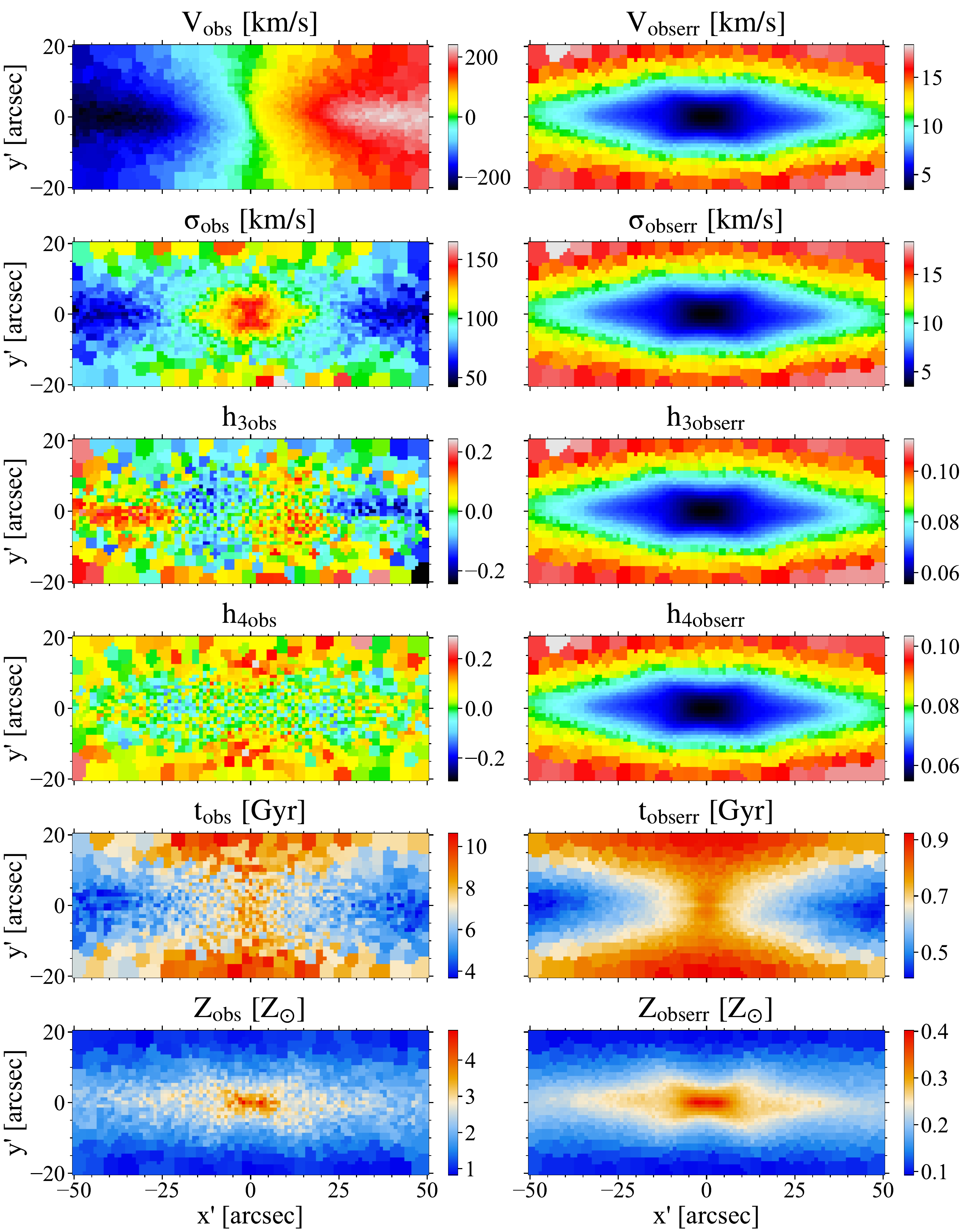}
    \caption{Mock stellar kinematic, age, and metallicity maps (left panels), and their corresponding error maps (right panels) for Au-23 with viewing angles, $(\theta_{\rm T},\varphi_{\rm T})=(85^\circ,50^\circ),$ with $\rm 1\,arcsec=0.2\,kpc$. From top to bottom: Mean velocity, $V$; velocity dispersion, $\sigma$; third-order, $h_3$, and fourth-order, $h_4$, Gauss-Hermite coefficients; stellar age, $t$; and stellar metallicity, $Z$. The kinematic maps and their error maps are the same as Fig. 2 of \citet{Jin2025}. We note that the kinematic error maps exhibit similar statistical properties as they are generated from particle noise \citep{Tsatsi2015}, but their ranges are different (as indicated by the colour bars in the right panels).}
    \label{observational-data}
\end{figure*}

The Auriga project is a suite of cosmological magnetohydrodynamical zoom-in simulations that consists of 30 isolated Milky Way-mass halos \citep{Grand2017}, which are selected from the dark matter-only version of the EAGLE Ref-L100N1504 simulation \citep{Schaye2015}. We adopted three simulated galaxies from the Auriga simulations: Au-18, Au-23, and Au-28, which are the same as in our previous paper \citep{Jin2025}. Au-18 and Au-23 have clear BP/X-shaped structures, while Au-28 does not. The stellar masses of these three galaxies are $M_*=8.04\times10^{10}\,\rm M_\odot$, $9.02\times10^{10}\,\rm M_\odot$, and $10.45\times10^{10}\,\rm M_\odot$, respectively, and their corresponding dark matter halo masses are $M_{\rm 200}=1.22\times10^{12}\,\rm M_\odot$, $1.58\times10^{12}\,\rm M_\odot$, and $1.61\times10^{12}\,\rm M_\odot$.

For the three simulated galaxies, we adopted the mock surface brightness and stellar kinematic maps created in \citet{Jin2025}. Each simulated galaxy was observed with four different projections $(\theta_{\rm T},\varphi_{\rm T})=(85^\circ,50^\circ)$, $(85^\circ,80^\circ)$, $(89^\circ,50^\circ)$, and $(89^\circ,80^\circ)$, where $\theta_{\rm T}$ is the inclination angle and $\varphi_{\rm T}$ is the bar azimuthal angle ($\varphi_{\rm T}=0^\circ$ represents that the bar is perfectly end-on, while $\varphi_{\rm T}=90^\circ$ denotes that the bar is perfectly side-on). The redshift $z=0$ snapshots were adopted and the galaxies were placed at a distance of $\rm41.25\,Mpc$ ($\rm 1\,arcsec=0.2\,kpc$), which is similar to the median distance of the GECKOS sample ($15<D\,\rm[Mpc]<70$; \citealp{vdSande2024}).

The mock surface brightness was constructed from the projected stellar mass distribution by assuming a constant stellar mass-to-light ratio of $M_*/L=2$. We note that different values of the constant $M_*/L$ do not affect the results, as luminosity is treated as a scale factor in our models. The surface brightness maps were initially constructed with a pixel size of $\rm 1\times1\,arcsec^2$. These pixels were then adaptively binned using the Voronoi binning method \citep{Cappellari2003} to generate binned apertures that match the typical scale of those in MUSE kinematic maps at $S/N=50$ to 100 (despite our initial pixels being coarser than MUSE's $\rm 0.2\times0.2\,arcsec^2$ pixels). The mock kinematic maps were derived by fitting the velocity distribution of the stellar particles in each aperture with a Gauss-Hermite expansion \citep{Gerhard1993,vdMarel1993,Rix1997} and were then perturbed by adding Gaussian noise based on particle numbers following the method in \citet{Tsatsi2015}. We refer to \citet{Jin2025} for the details of creating these mock data. We have 12 mock datasets (three simulated galaxies, each with four different projections) in total. Each mock dataset is labelled with the format `name-$\theta_{\rm T}$-$\varphi_{\rm T}$'; for example, the mock observation of Au-23 with viewing angles $(\theta_{\rm T},\varphi_{\rm T})=(85^\circ,50^\circ)$ is denoted as Au-23-85-50. This notation is used consistently throughout the paper.

We further created mock stellar age maps and stellar metallicity maps for each projection $(\theta_{\rm T},\varphi_{\rm T})$ of each galaxy. For both the age and metallicity maps, we used the same apertures as the kinematic maps. Within each aperture, $i$, we determined the mean age, $t_i$, and the mean metallicity, $Z_i/Z_{\odot}$, by calculating the mass-weighted values of all stellar particles located within the aperture. We assumed the percentage errors of the age and metallicity maps to be $10\%$ and added Gaussian perturbations to the maps: $t_i'=t_i+0.1t_ia$, and $Z_i'=Z_i+0.1Z_ib$, where $a$ and $b$ are random values with standard deviations equal to unity. The perturbed age maps, $t_i'$, and metallicity maps, $Z_i'/Z_{\odot}$, were taken as mock observations, while $0.1t_i$ and $0.1Z_i$ were treated as corresponding error maps. Fig.~\ref{observational-data} illustrates the stellar kinematics, ages, and metallicities for Au-23-85-50.

\section{Barred population-orbit superposition models}
\label{sec3}
\begin{figure*}
    \centering
    \includegraphics[width=14cm]{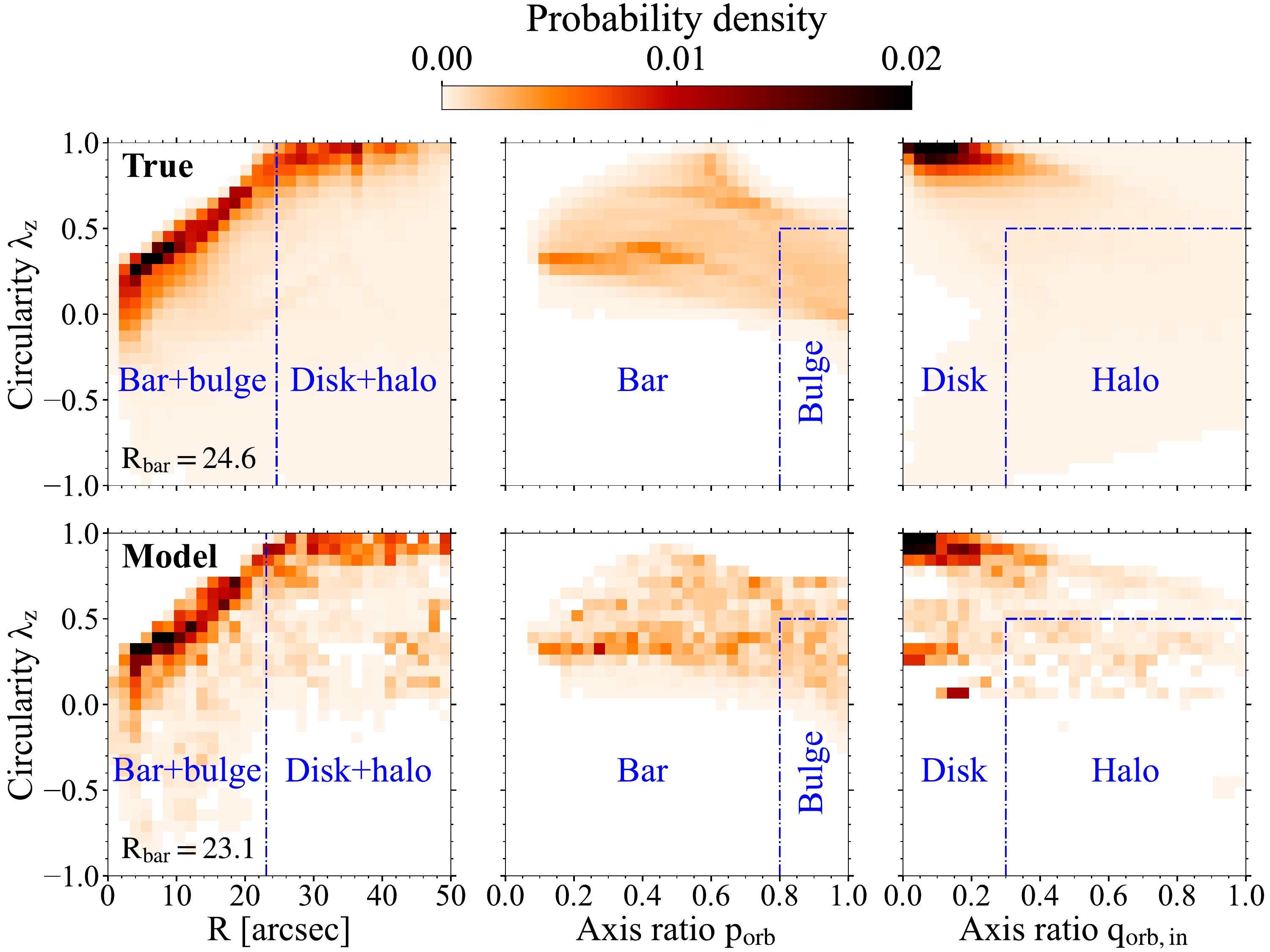}
    \caption{Structural decomposition based on orbital properties demonstrated using Au-23-85-50 and compared with the truth. Four properties are used to decompose the galaxy: Circularity, $\lambda_z$; time-averaged radius, $R$; axis ratio, $p_{\rm orb}$, in the $x$-$y$ plane; and axis ratio, $q_{\rm orb}$, in the $x$-$z$ plane. The top panels display the true distributions while the bottom panels correspond to the model results. Left panels: Stellar orbit distributions of the entire galaxy in the $\lambda_z$--$R$ phase space. We calculate the $\rm1\,kpc$ moving average of the cold orbit fraction $f_{\rm cold}$ ($\lambda_z\ge0.8$; $\rm 1\,arcsec=0.2\,kpc$) and define the dynamical bar length, $R_{\rm bar}$, as the smallest radius where $f_{\rm cold}\ge0.5$. Orbits with $R\le R_{\rm bar}$ are classified as bar+bulge components; those with $R>R_{\rm bar}$ are categorised as disc+halo components. The values of $R_{\rm bar}$ are indicated by the blue dashed lines and the annotations. Middle panels: Stellar orbit distributions for bar and bulge components in the $\lambda_z$--$p_{\rm orb}$ phase space. Orbits with $\lambda_z\le0.5$ and $p_{\rm orb}\ge0.8$ are assigned to the bulge; other orbits constitute the bar. The boundaries between the bar and bulge are shown by the blue dashed lines. Right panels: Stellar orbit distributions for disc+halo components in the $\lambda_z$--$q_{\rm orb,in}$ phase space. Orbits with $\lambda_z\le0.5$ and $q_{\rm orb,in}\ge0.3$ are categorised as the stellar halo; while the other orbits make up the disc. The boundaries between the disc and halo are shown by the blue dashed lines.}
    \label{orbit-distributions-combined}
\end{figure*}
\begin{figure*}
    \centering
    \includegraphics[width=17.8cm]{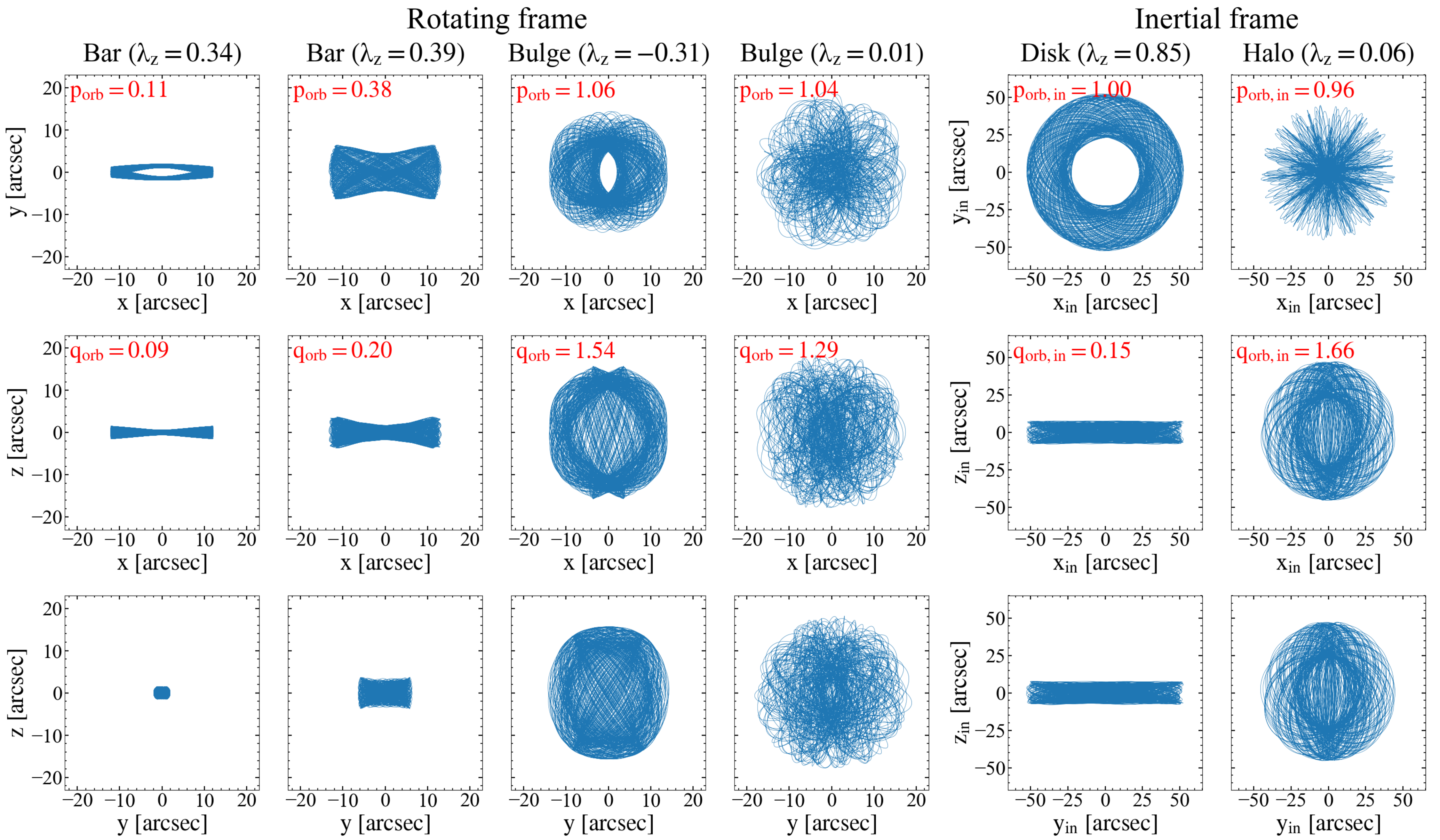}
    \caption{Trajectories of several representative orbits for Au-23-85-50. The first four columns display the orbital trajectories in the $x$-$y$, $x$-$z$, and $y$-$z$ planes (the bar's rotating frame), including two bar orbits and two bulge orbits, with their axis ratios $p_{\rm orb}$ and $q_{\rm orb}$ presented in the text. The fifth and sixth columns show the orbital trajectories in the $x_{\rm in}$-$y_{\rm in}$, $x_{\rm in}$-$z_{\rm in}$, and $y_{\rm in}$-$z_{\rm in}$ planes (the inertial frame), including a disc orbit and a halo orbit, with their axis ratios, $p_{\rm orb,in}$ and $q_{\rm orb,in}$, presented in the text. We note that $p_{\rm orb}$ and $q_{\rm orb}$ are calculated in the bar's rotating frame, while $\lambda_z$, $p_{\rm orb,in}$, and $q_{\rm orb,in}$ are calculated in the inertial frame.}
    \label{orbit-trajectories}
\end{figure*}
\begin{figure*}
    \centering
    \includegraphics[width=17.8cm]{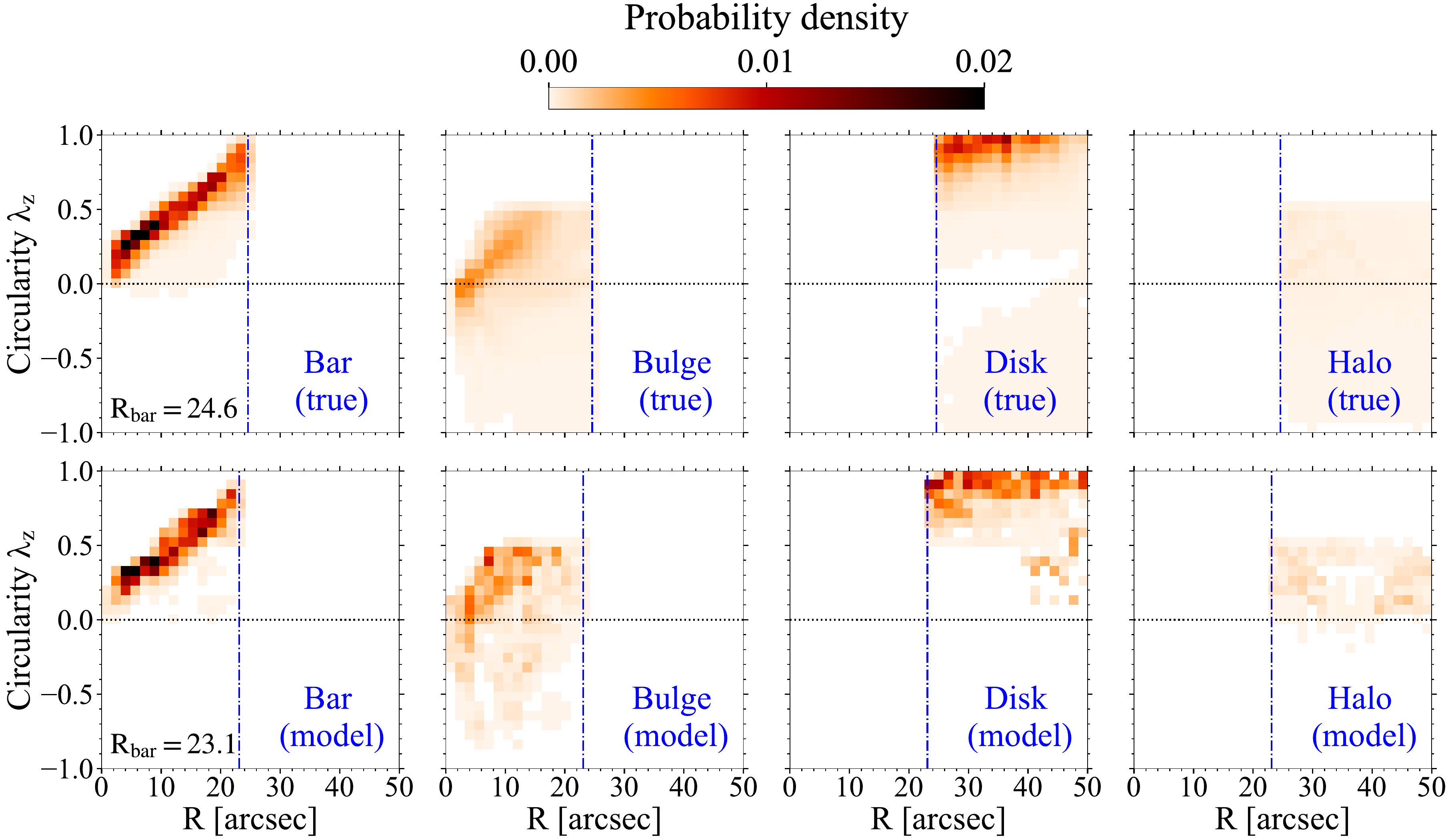}
    \caption{True and model-predicted probability density distributions of stellar orbits in the $\lambda_z$--$R$ phase space for different structures in Au-23. The top panels indicate the true distributions while the bottom panels represent the model results from Au-23-85-50. From left to right, the panels show the stellar orbit distributions for the bar, the bulge, the disc, and the stellar halo. The probability densities of all orbits within the data coverage ($R\le\rm50\,arcsec$) are normalised to unity, with their values indicated by the colour bar. The horizontal black dotted lines denote $\lambda_z=0$. The vertical blue dashed lines represent the true and model-predicted bar lengths, $R_{\rm bar}$, derived from orbit analysis, with their values shown in the text.}
    \label{orbit-distributions}
\end{figure*}
\begin{figure*}
    \centering
    \includegraphics[width=14cm]{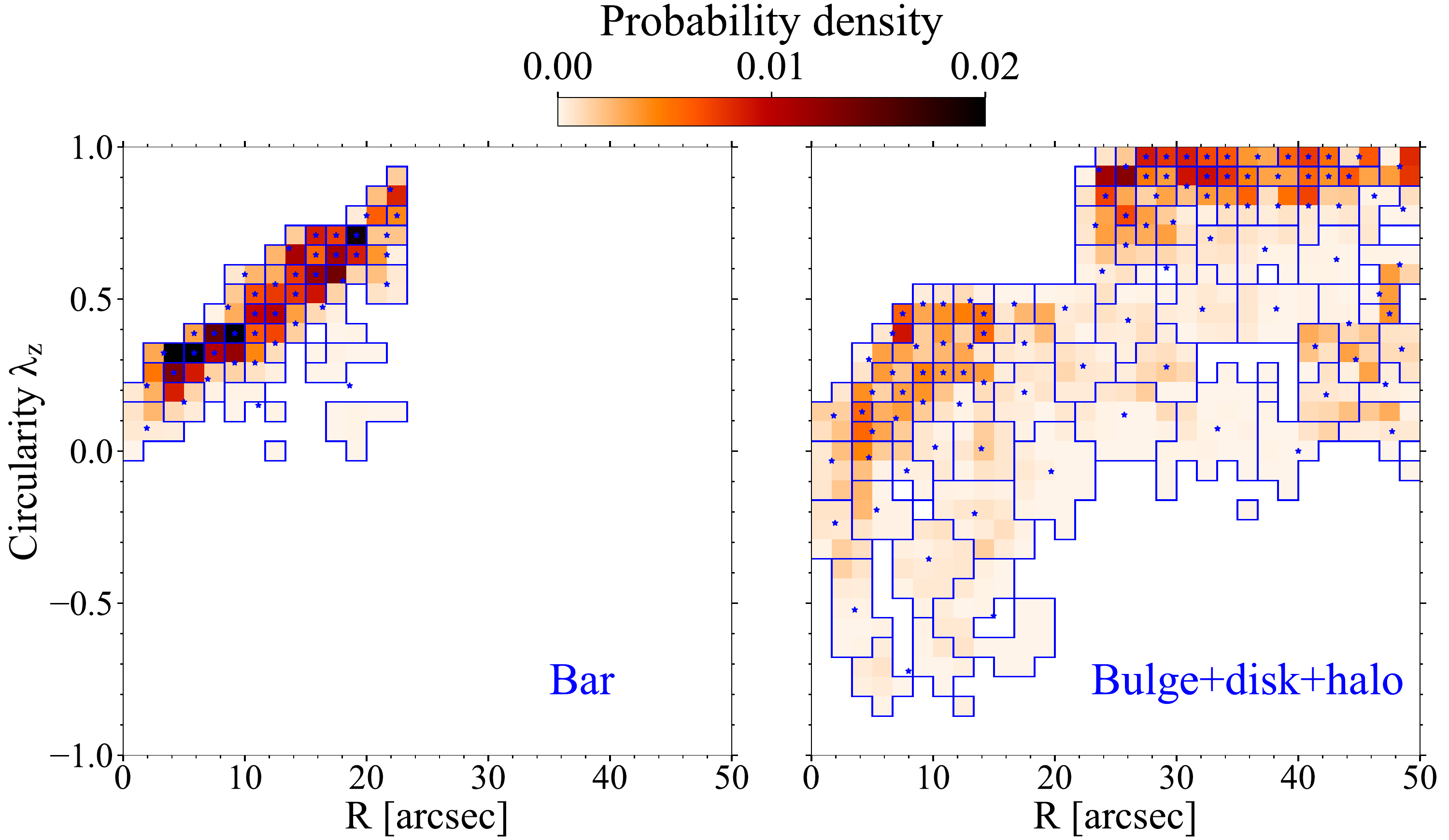}
    \caption{ Orbit bundles divided in the phase space of circularity, $\lambda_z$, versus time-averaged radius, $R$, for Au-23-85-50. The left panel represents bar orbits while the right panel is for orbits from other components, including the bulge, disc, and stellar halo. The probability densities of all orbits within the data coverage ($R\le\rm50\,arcsec$) are normalised to unity, with their values indicated by the colour bar. For each panel, orbits in $\lambda_z$--$R$ phase space are divided into different bundles by using the Voronoi binning method, with each bundle containing $\gtrsim0.5\%$ of the total orbit weight. The blue asterisks and lines indicate the centres and boundaries of Voronoi bins, respectively.}
    \label{Voronoi-bins}
\end{figure*}
\begin{figure*}
    \centering
    \includegraphics[width=16cm]{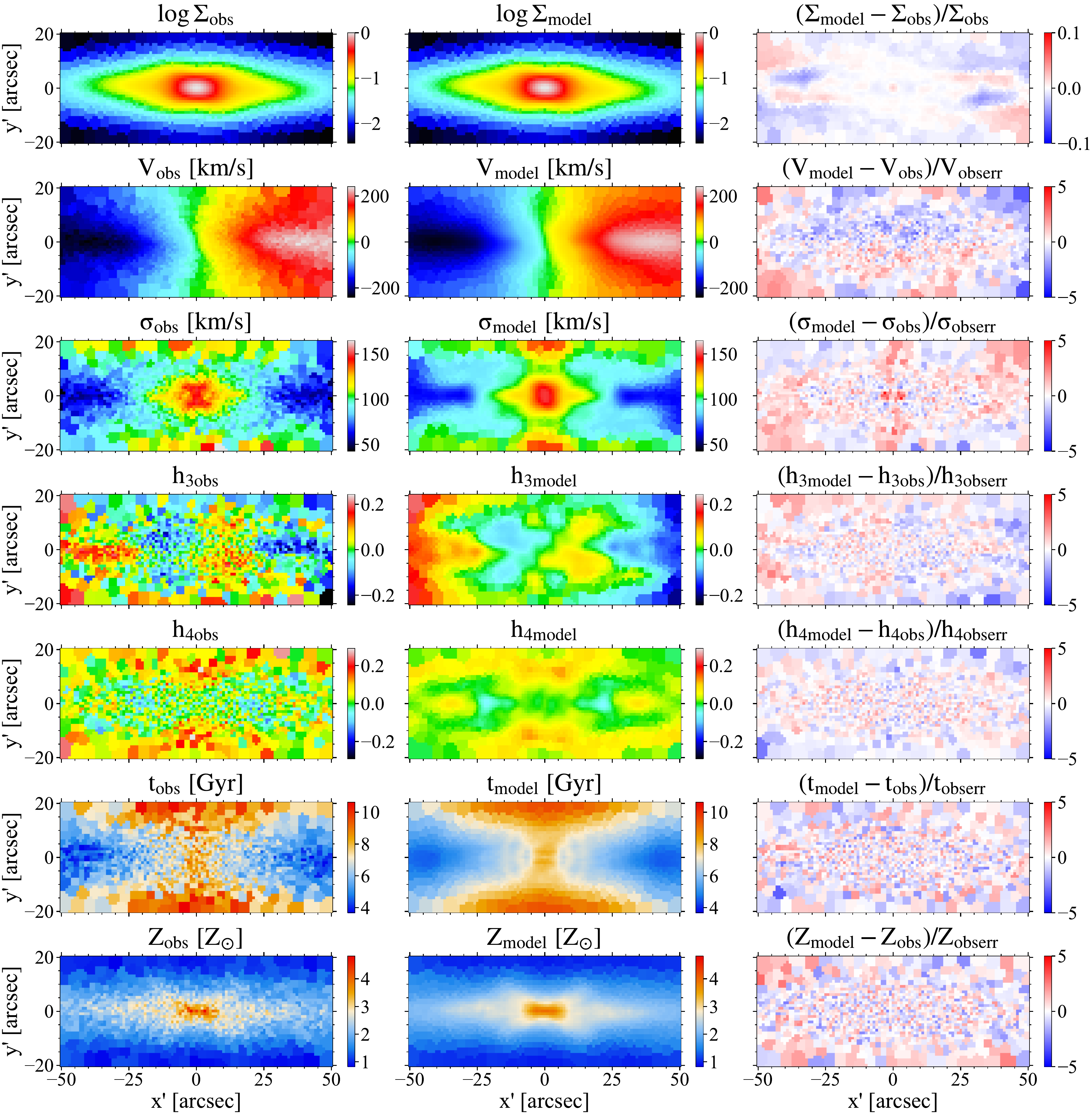}
    \caption{Mock surface brightness, stellar kinematic, age, and metallicity maps (left panels), best-fitting model for Au-23-85-50 (middle panels), and residuals (right panels). From top to bottom: Normalised surface brightness, $\log\Sigma$; mean velocity, $V$; velocity dispersion, $\sigma$; third-order, $h_3$, and fourth-order, $h_4$, Gauss-Hermite coefficients; stellar age, $t$; and stellar metallicity, $Z$. The residuals for surface brightness are relative deviations, while those for kinematics and stellar populations are standardised residuals. The surface brightness and kinematic maps are the same as Fig. 6 of \citet{Jin2025}.}
    \label{best-fitting-maps}
\end{figure*}

We propose a three-step methodology to construct barred population-orbit superposition models. First, we create barred orbit-superposition models by simultaneously fitting the stellar luminosity distributions and kinematic data, and search for the best-fitting model (Sect.~\ref{sec3.1}). This step builds directly on our previous work \citep{Jin2025}, which contains detailed explanations and so, we only briefly summarise it here. Then, we categorise the stellar orbits of the best-fitting model into different components based on their orbital properties (Sect.~\ref{sec3.2}). Finally, we tag the orbits with ages and metallicities by fitting the stellar age and metallicity maps (Sect.~\ref{sec3.3}).

\subsection{Barred orbit-superposition models}
\label{sec3.1}
The procedures for creating a barred orbit-superposition model include three major steps: (1) constructing the gravitational potential using a set of free parameters; (2) sampling and integrating orbits; and (3) solving orbit weights by fitting the observational data. By exploring the free parameter space, the above steps are repeated until the best-fitting model is found.

The gravitational potential of a barred galaxy contains three components: (1) a triaxial stellar bar with constant figure rotation, $\rm\Omega_p$; (2) an axisymmetric stellar disc with its major axis aligned with the bar; and (3) a spherical dark matter halo. The central black hole is not considered in the modelling due to the limited spatial resolution ($\rm\sim0.2\,kpc/pixel$) of the mock data. To calculate the stellar potential, the galaxy's surface brightness is first fitted by the multi-Gaussian expansion (MGE) formalism \citep{Emsellem1994,Cappellari2002}, with each Gaussian being assigned either to the triaxial bar component or to the axisymmetric disc component. The 2D Gaussian components of the bar and disc are separately deprojected into the 3D luminosity density using distinct viewing angles: the inclination angle, $\theta_{\rm disc}$, and the position angle, $\psi_{\rm disc}$, for the axisymmetric disc and the inclination angle, $\theta_{\rm bar}$, the bar azimuthal angle, $\varphi$, and the position angle, $\psi_{\rm bar}$, for the triaxial bar. Taking the assumption that the bar major axis ($x$-axis) is in the disc plane, the inclination angles of both components become equal ($\theta_{\rm bar}=\theta_{\rm disc}=\theta$). The disc in the observing plane is rotated to make sure its major axis aligns with the $x'$-axis ($\psi_{\rm disc}=90^\circ$). For edge-on views, the bar position angle, $\psi_{\rm bar}$, which represents the misalignment between the bar major axis and the disc major axis in the observing plane, is difficult to measure directly. However, small changes in $\psi_{\rm bar}$ will significantly alter the morphology of the deprojected bar. Therefore, $\psi_{\rm bar}$ is determined in each set of ($\theta,\varphi$) by adding physical constraints to the bar morphology; thus, only the viewing angles $\theta$ and $\varphi$ remain free during the deprojection (see Sect. 3.1.1 in \citealp{Jin2025} for details). By multiplying the 3D luminosity density by a constant mass-to-light ratio, $M_*/L$, and solving Poisson's equation, both the potentials of the stellar bar and the stellar disc can be derived. We note that the bar and disc used here to construct the stellar potential are distinct from the dynamically decomposed bar and disc mentioned later (see Sect.~\ref{sec3.2}). The spherical dark matter potential is described by a three-parameter generalised NFW (gNFW) profile (\citealp{Navarro1996,Zhao1996}; see also \citealp{Barnabe2012,Cappellari2013}). Overall, there are seven free parameters in the modelling: the inclination angle, $\theta$; the bar azimuthal angle, $\varphi$; the stellar mass-to-light ratio, $M_*/L$; the bar pattern speed, $\rm\Omega_p$; the dark matter concentration, $c$; the virial mass, $M_{200}$; and the inner density slope of dark matter, $\gamma$.

The initial conditions of the orbits are sampled in the $x$-$z$ plane of a Cartesian coordinate system, including a prograde orbit library and a retrograde orbit library. These orbits are then integrated within a gravitational potential defined by free parameters. This can produce common orbit types in barred galaxies, such as box orbits dominated by random motion, rotation-dominated $z$-tube orbits, and $x_1$ orbits that are always elongated along the bar.

The orbit weights are solved using the non-negative least squares (NNLS; \citealp{Lawson1974}) method, taking the 2D surface brightness fitted by MGE, the 3D luminosity density deprojected from MGE, and the kinematic data as constraints. After solving the orbit weights, the goodness of the model is evaluated by calculating the difference between the model-fitted and observed kinematic maps, which is expressed as
\begin{equation}
\begin{split}
    \chi^2=\sum_{i=1}^{N_{\rm obs}}\left[ \left(\frac{V^*_i-V_i}{\Delta V_i}\right)^2+\left(\frac{\sigma^*_i-\sigma_i}{\Delta \sigma_i} \right)^2+\right.&\\
    \left.\left(\frac{h^*_{3i}-h_{3i}}{\Delta h_{3i}} \right)^2+\left(\frac{h^*_{4i}-h_{4i}}{\Delta h_{4i}} \right)^2\right].\\
\end{split}
\label{chi2kin}
\end{equation}
The variables marked with a `$*$' indicate the model fittings, those with a `$\Delta$' represent the errors of the input data, and those without a marker denote the input data.

The best-fitting model is determined by searching the free parameter space through an iterative process. This iterative process starts with making initial guesses about the free parameters ($\theta,\varphi,M_*/L,\Omega_{\rm p},c,M_{200},\gamma$) and constructing initial models. Models with a relatively lower $\chi^2$ are selected, and new models are created by walking a few steps in the parameter space around the selected models. This process is repeated until a minimum $\chi^2$ is obtained, with all surrounding models calculated in the parameter space. The model with the minimum chi-square $\chi^2_{\rm min}$ is defined as the best-fitting model and is taken as the default model in our analysis.

\subsection{Decomposing galaxies based on stellar orbits}
\label{sec3.2}
The best-fitting model contains an ensemble of stellar orbits. In this paper, orbits are categorised into four distinct components: bar, bulge, disc, and stellar halo, based on four orbital properties, including (1) circularity, $\lambda_z$; (2) time-averaged radius, $R$; (3) axis ratio, $p_{\rm orb}$, in the $x$-$y$ plane (bar's rotating frame); and (4) axis ratio, $q_{\rm orb,in}$, in the $x_{\rm in}$-$z_{\rm in}$ plane (inertial frame). We note that circularity is defined exclusively in the inertial frame ($x_{\rm in},y_{\rm in},z_{\rm in}$), while the axis ratios are calculated in both the inertial frame and the bar's rotating frame ($x,y,z$).

The circularity represents the orbit's angular momentum, which is defined as a ratio of time-averaged quantities,
\begin{equation}
    \lambda_z=\overline{L_z}/(R\times\overline{V_{\rm rms}}),
\end{equation}
where $\overline{L_z}=\overline{x_{\rm in}v_{y\rm in}-y_{\rm in}v_{x\rm in}}$, $R=\overline{\sqrt{x_{\rm in}^2+y_{\rm in}^2+z_{\rm in}^2}}$ and $\overline{V_{\rm rms}}=\sqrt{\overline{v_{x\rm in}^2+v_{y\rm in}^2+v_{z\rm in}^2+2v_{x\rm in}v_{y\rm in}+2v_{x\rm in}v_{z\rm in}+2v_{y\rm in}v_{z\rm in}}}$.
Orbits with $\lambda_z\sim1$ are rotation-dominated, orbits with $\lambda_z\sim0$ are dispersion-dominated, and orbits with $\lambda_z<0$ correspond to retrograde orbits.

Orbit circularity, $\lambda_z$, and radius, $R,$ have been widely used in previous studies to decompose galaxies without considering the bars (e.g. \citealp{Zhu2018a,Zhu2018b,Zhu2018c,Zhu2020,Zhu2022a,Zhu2022b,Zhu2025,Jin2019,Jin2020,Jin2024,Poci2019,Poci2021,Santucci2022,Santucci2023,Santucci2024,Ding2023,Ding2024,Zhang2025}).
However, for barred galaxies, these criteria alone fail to separate the bulge and bar in the innermost regions, as both structures exhibit orbits with $\lambda_z\sim0$. Additionally, in the outer regions of galaxies, there are non-circular disc orbits with obvious radial motion ($\lambda_z\ll1$), which overlap with the circularity range of halo orbits.

To break the degeneracy between different structures in the $\lambda_z$--$R$ phase space, we further introduce the axis ratios, $p_{\rm orb}=b/a$ and $q_{\rm orb}=c/a$, calculated from orbital trajectories, with $a$, $b$, and $c$ denoting the time-averaged amplitudes along the $x$-axis, $y$-axis, and $z$-axis, respectively. Thus, $p_{\rm orb}$ and $q_{\rm orb}$ are expressed as
\begin{equation}
    p_{\rm orb}=\frac{\overline{|y-\overline{y}|}}{\overline{|x-\overline{x}|}},\quad
    q_{\rm orb}=\frac{\overline{|z-\overline{z}|}}{\overline{|x-\overline{x}|}}.
\end{equation}
Similarly, we calculated the axis ratios, $p_{\rm orb,in}$ and $q_{\rm orb,in}$, of the orbits in the inertial frame ($x_{\rm in},y_{\rm in},z_{\rm in}$).

From the kinematic perspective, the disc is dominated by dynamically cold orbits ($\lambda_z\ge0.8$), whereas the bar rotates as a rigid body with low $\lambda_z$ in its inner region and becomes corotating with the disc at the bar end. Therefore, we defined the dynamical bar length, $R_{\rm bar}$, as the radius where cold orbits begin to dominate. We calculated the $\rm1\,kpc$ moving average of the cold orbit fraction, $f_{\rm cold}$, and identified $R_{\rm bar}$ as the smallest radius where $f_{\rm cold}\ge0.5$. In this way, we obtained $R_{\rm bar,model}=\rm23.1\,arcsec$ for the model of Au-23-85-50.

For each simulated galaxy, we froze all the stellar and dark matter particles in the redshift $z=0$ snapshot and calculated the gravitational potential from their distributions using multipole expansion of spherical harmonics \citep{Binney2008}. Taking the 6D phase-space information of stellar particles as the initial conditions, we integrated orbits for $\sim 100$ orbital periods in the bar's rotating frame (with true pattern speeds from \citealp{Fragkoudi2020,Fragkoudi2021}) via the AGAMA software \citep{Vasiliev2019}\footnote{\url{https://github.com/GalacticDynamics-Oxford/Agama}}. We then calculated the same parameters characterising the orbits. The true bar length, $R_{\rm bar}$, was then determined the same way as in the model and we obtained $R_{\rm bar,true}=\rm24.6\,arcsec$ for Au-23.

We identified different structures based on orbital information using an identical methodology in both our models and the simulations. We first divided the orbits with $R\le R_{\rm bar}$ as the bar+bulge components and those with $R>R_{\rm bar}$ as the disc+halo components, using $R_{\rm bar,model}$ and $R_{\rm bar,true}$ for our models and the simulations, respectively. Then we further classified the different components according to the following criteria: (1) bar orbits are rotation-dominated ($\lambda_z>0.5$) and/or elongated along the bar ($p_{\rm orb}<0.8$); (2) bulge orbits are random-motion-dominated ($\lambda_z\le0.5$) and not elongated along the bar ($p_{\rm orb}\ge0.8$); (3) disc orbits are rotation-dominated ($\lambda_z>0.5$) and/or thin ($q_{\rm orb,in}<0.3$); and (4) halo orbits are random-motion-dominated ($\lambda_z\le0.5$) and thick ($q_{\rm orb,in}\ge0.3$). In Fig.~\ref{orbit-distributions-combined}, we exemplify these criteria using Au-23-85-50. The trajectories of several representative orbits belonging to different structures for Au-23-85-50 are illustrated in Fig.~\ref{orbit-trajectories}.

We show the probability density distributions of stellar orbits for Au-23 and those recovered by the model of Au-23-85-50 in Fig.~\ref{orbit-distributions}. We successfully decomposed the galaxy into a rigidly rotating bar ($\lambda_z$ is strongly correlated with $R$), a random-motion-dominated bulge ($|\lambda_z|\lesssim0.5$), a rotation-dominated disc ($\lambda_z\gtrsim0.8$), and a stellar halo; the stellar orbit distributions of these components are similar in the true simulation and our model.

\subsection{Tagging the stellar orbits with ages and metallicities}
\label{sec3.3}
After decomposing the galaxy, we separated the bar orbits and bulge, disc, and halo orbits into two distinct $\lambda_z$--$R$ phase spaces. Following \citet{Zhu2020} and \citet{Jin2024}, we then used the Voronoi 2D binning method \citep{Cappellari2003} to divide orbits in each $\lambda_z$--$R$ phase space into different orbit bundles, with each bundle containing $\gtrsim0.5\%$ of the total orbit weight. This created 100-200 bundles in total. Fig.~\ref{Voronoi-bins} illustrates this procedure for Au-23-85-50.

The stellar orbits are tagged with ages and metallicities based on the assumption that stars on similar orbits have similar stellar populations. Thus, we assume that each orbit bundle $k$ has a simple stellar population with age $t_k$ and metallicity $Z_k$ to be determined. By projecting the orbits, the mean age and metallicity of the aperture $i$ in the observing plane can be written as
\begin{equation}
    t_{\rm model}^i=\frac{\sum_{k=1}^{N_b} t_k f_k^i}{\sum_{k=1}^{N_b} f_k^i},\quad
    Z_{\rm model}^i=\frac{\sum_{k=1}^{N_b} Z_k f_k^i}{\sum_{k=1}^{N_b} f_k^i},
\end{equation}
where $N_b$ is the total number of orbit bundles, and $f_k^i$ is the luminosity contributed by the orbit bundle, $k$, in aperture, $i$. By matching the model-fitted maps with the observed data, the age, $t_k$, and metallicity, $Z_k$, of each orbit bundle can be determined.

Similarly to the kinematics, the goodness of model-fitted age and metallicity maps is evaluated by their chi-square difference from observations, which are expressed as
\begin{equation}
    \chi_{\rm age}^2=\sum_{i=1}^{N_{\rm obs}}\left(\frac{t_{\rm obs}^i-t_{\rm model}^i}{t_{\rm obserr}^i}\right)^2,\quad
    \chi_{\rm met}^2=\sum_{i=1}^{N_{\rm obs}}\left(\frac{Z_{\rm obs}^i-Z_{\rm model}^i}{Z_{\rm obserr}^i}\right)^2.
\end{equation}

From the age and metallicity maps, the youngest outer disc's age and the metal-poorest halo's metallicity can be directly predicted. However, ages and metallicities observed in the central regions are mixtures of different stellar populations, revealing the possible presence of intrinsically more metal-rich and older stellar populations than those observed from the 2D maps. Thus, we set the age and metallicity ranges of orbits as
\begin{equation}
    \min(t_{\rm obs}^i)\le t_k\le \min[\max(t_{\rm obs}^i)+\Delta(t_{\rm obs}^i), \rm14\,Gyr],
\end{equation}
\begin{equation}
    \min(Z_{\rm obs}^i)\le Z_k\le \max(Z_{\rm obs}^i)+\Delta(Z_{\rm obs}^i),
\end{equation}
where `$\Delta$' represents the standard deviation of the observed data.

We adopted the bounded-variable least squares method (implemented by Python Scipy\footnote{\url{https://scipy.org/}}), which is an optimisation algorithm that solves linear least-squares problems with bounded parameters, to minimise $\chi_{\rm age}^2$ and $\chi_{\rm met}^2$ individually. By solving the bounded-variable least squares, the ages and metallicities for all orbits can be derived and are taken as the best-fitting stellar population model. The results of the entire best-fitting model, including surface brightness, kinematics, ages, and metallicities, are shown in Fig.~\ref{best-fitting-maps}.

\section{Model-predicted structures and their true counterparts for an individual case}
\label{sec4}
\begin{figure*}
    \centering
    \includegraphics[width=17.8cm]{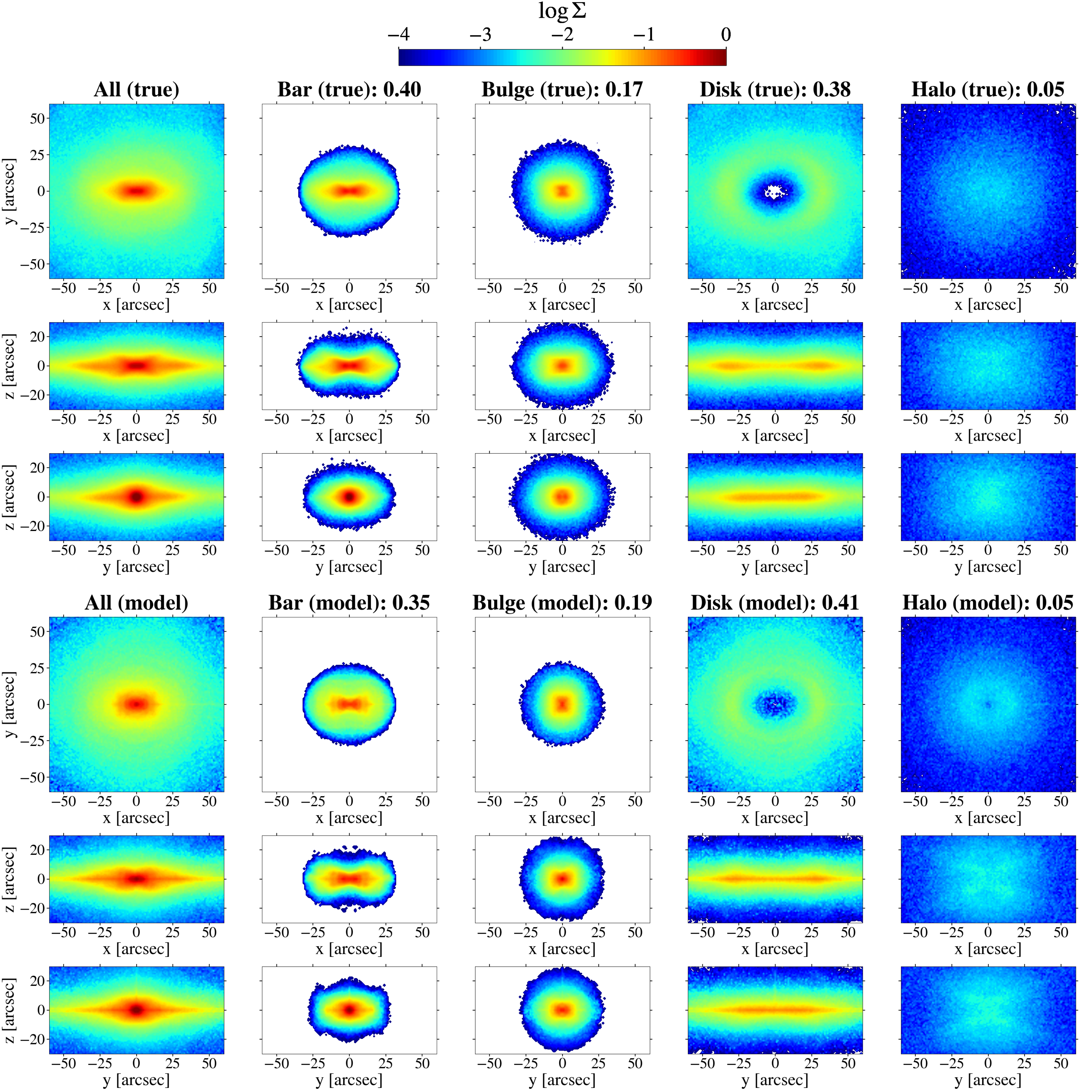}
    \caption{Spatial distributions of stellar mass for model-decomposed structures in Au-23-85-50 compared with their true counterparts. The top three rows display the true mass surface densities of Au-23 projected onto the internal $x$-$y$, $x$-$z$, and $y$-$z$ planes, while the bottom three rows show the model-predicted results for Au-23-85-50, with the central mass density normalised to unity. From left to right, the panels show the mass surface densities of the entire galaxy, the bar, the bulge, the disc, and the stellar halo. The mass fractions of all components were calculated within the $\rm60\times60\times30\,arcsec^3$ volume (aligned with the axis ranges in the panels), with their values shown in the text.}
    \label{mass-distributions}
\end{figure*}
\begin{figure*}
    \centering
    \includegraphics[width=17.8cm]{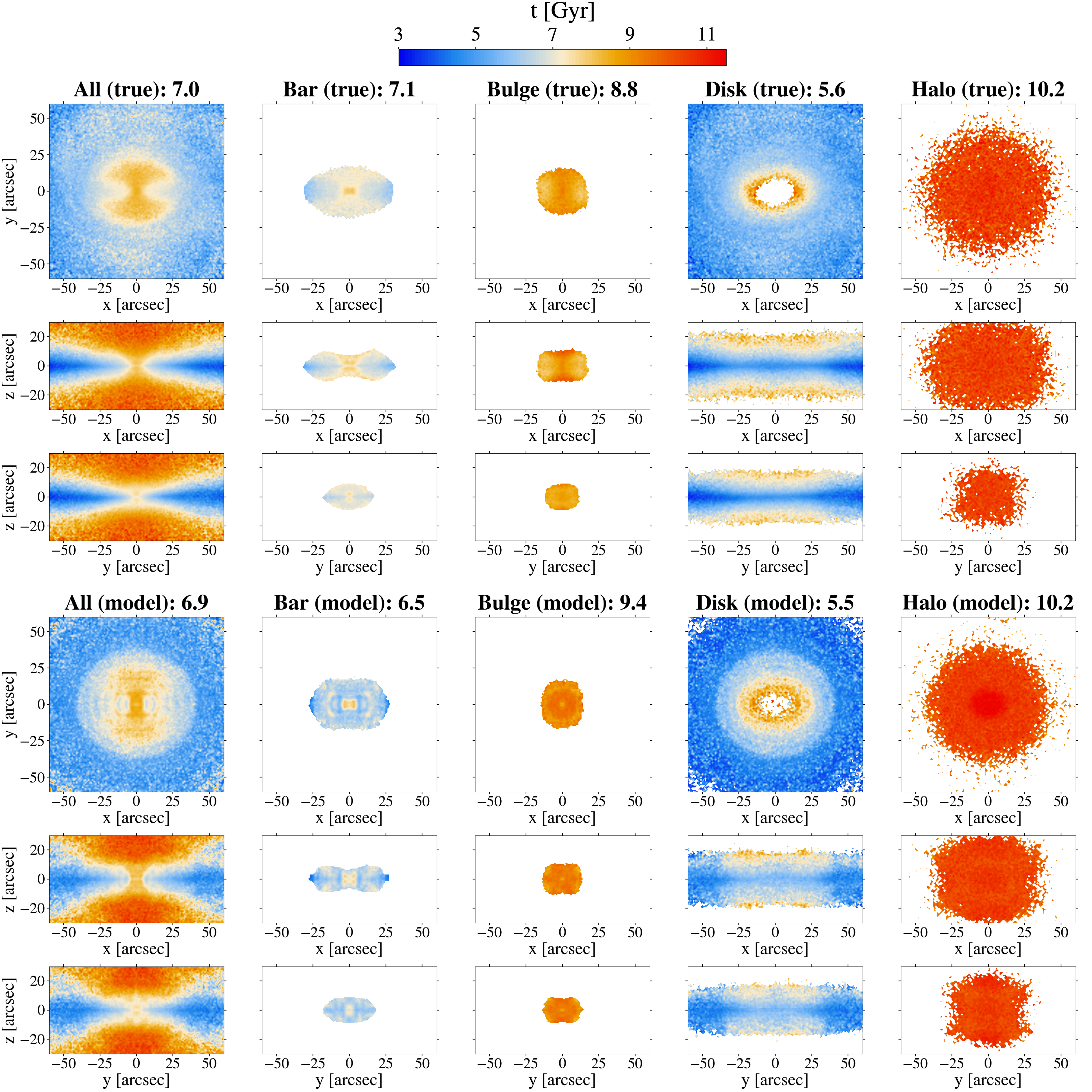}
    \caption{Spatial distributions of stellar ages for model-decomposed structures in Au-23-85-50 compared with their true counterparts. The top three rows display the stellar age distributions of Au-23 projected onto the internal $x$-$y$, $x$-$z$, and $y$-$z$ planes, while the bottom three rows show the model-predicted results for Au-23-85-50. From left to right, the panels show the stellar age distributions of the entire galaxy, the bar, the bulge, the disc, and the stellar halo. To highlight the age distributions, we only plot pixels brighter than $1\%$ of the galaxy's brightness pixel for the bar and bulge, and $0.1\%$ for the disc and halo. The mass-weighted mean stellar ages of the entire galaxy and all components were calculated within the $\rm60\times60\times30\,arcsec^3$ volume (aligned with the axis ranges in the panels), with their values shown in the text.}
    \label{age-distributions}
\end{figure*}
\begin{figure*}
    \centering
    \includegraphics[width=17.8cm]{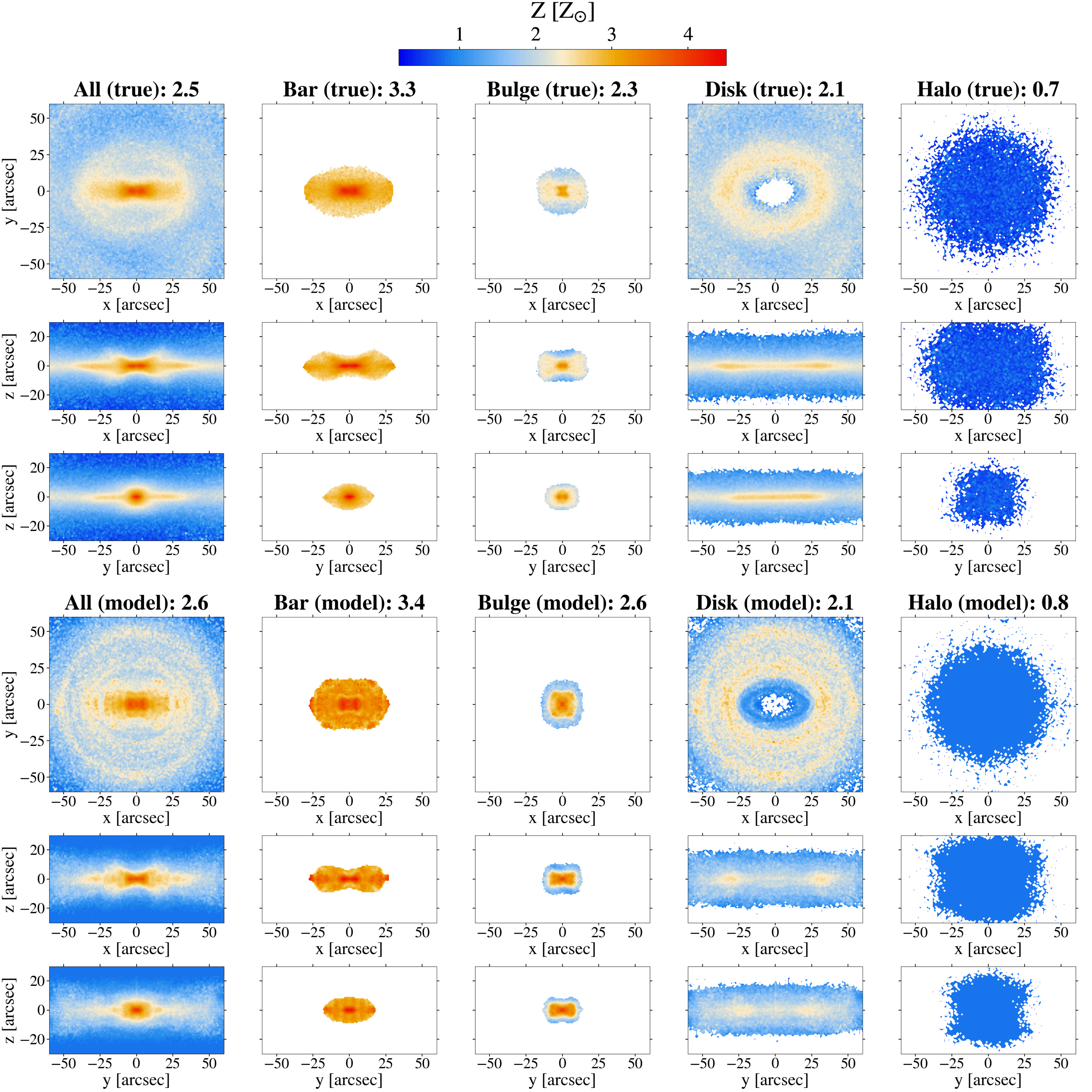}
    \caption{Spatial distributions of stellar metallicities for model-decomposed structures in Au-23-85-50 compared with their true counterparts. Each panel follows the same format as Fig.~\ref{age-distributions}.}
    \label{metallicity-distributions}
\end{figure*}
\begin{figure*}[!htb]
    \centering
    \includegraphics[width=16cm]{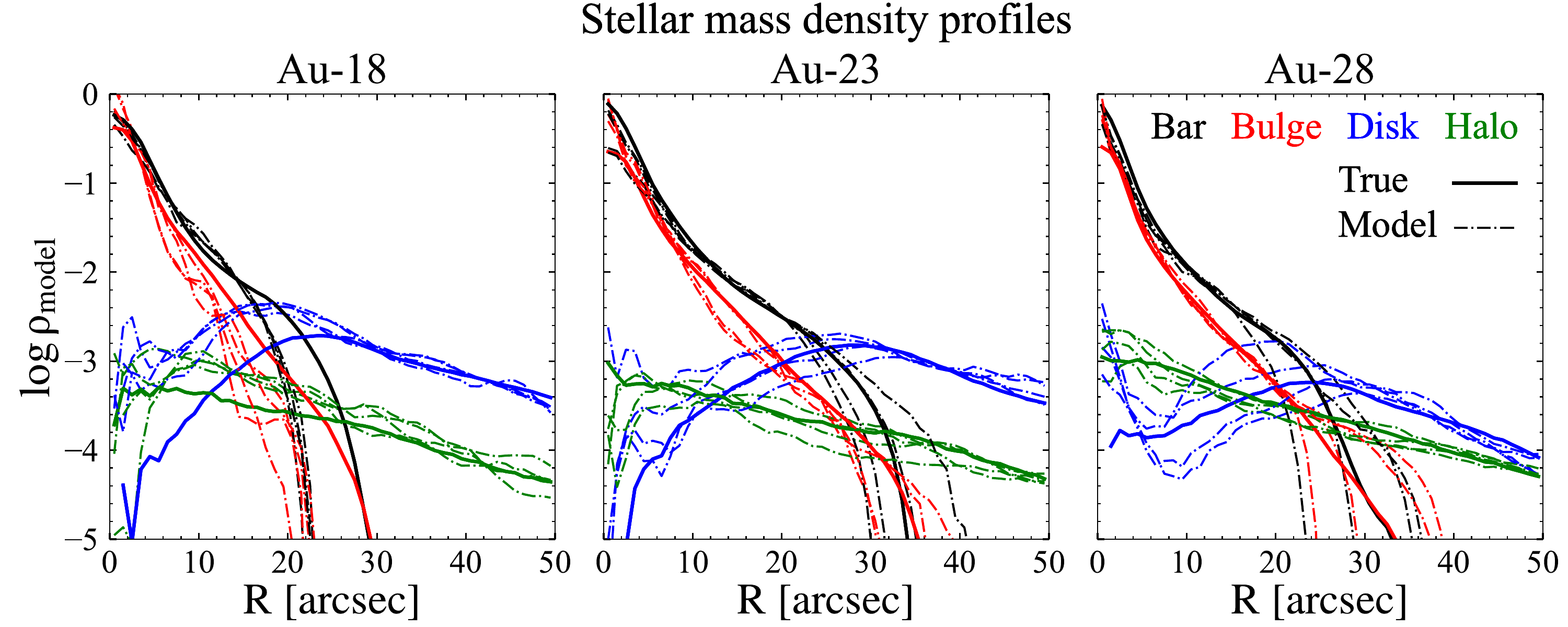}
    \caption{Comparison of the true and model-predicted mass density profiles of different structures. From left to right, the panels display normalised mass density profiles for Au-18, Au-23, and Au-28. The black, red, blue, and green lines denote the bar, bulge, disc, and stellar halo components, respectively. The solid curves represent the true mass density profiles, while the dashed lines denote model-predicted profiles derived from mock datasets with different viewing angles.}
    \label{mass-density-profiles}
\end{figure*}
\begin{figure*}[!htb]
    \centering
    \includegraphics[width=16cm]{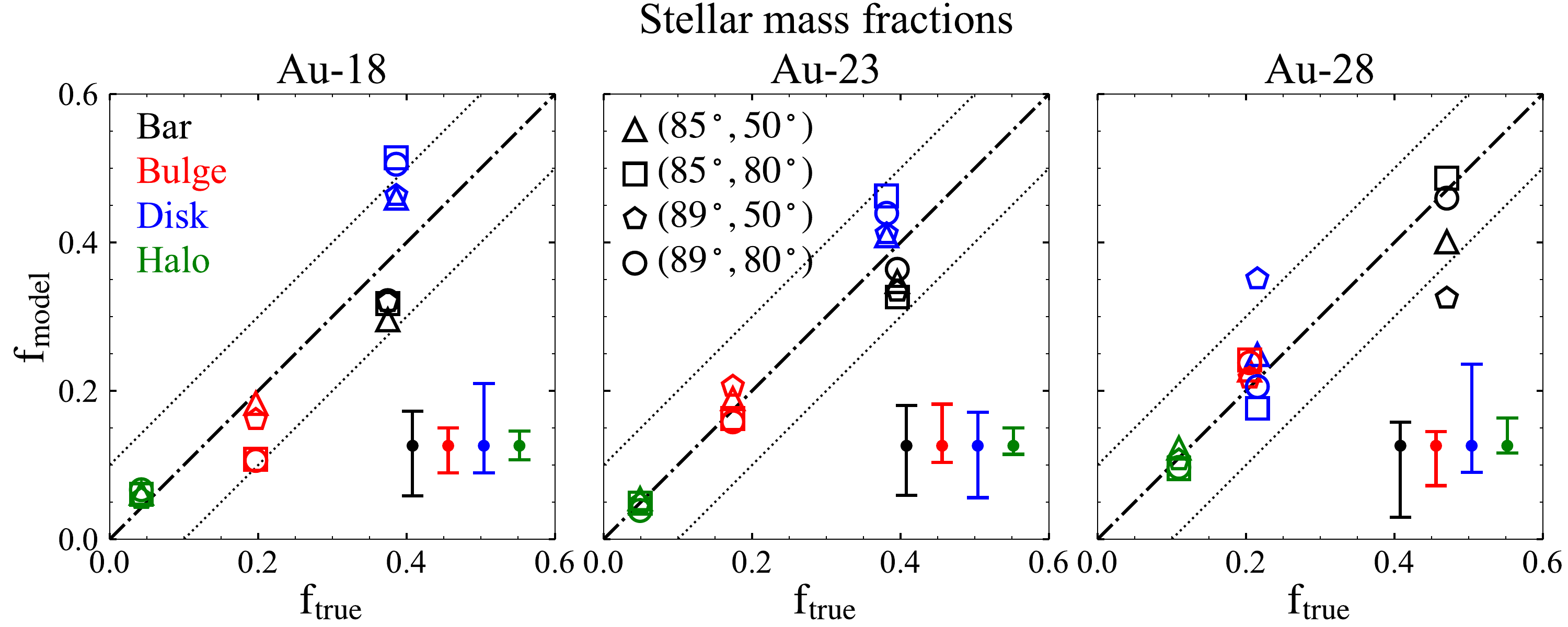}
    \caption{One-to-one comparison of the true and model-predicted mass fractions of decomposed structures. From left to right, the panels display mass fractions for Au-18, Au-23, and Au-28. The black, red, blue, and green markers denote the bar, bulge, disc, and stellar halo components, respectively. The triangles, squares, pentagons, and circles indicate mock datasets with viewing angles $(\theta_{\rm T},\varphi_{\rm T})=(85^\circ,50^\circ)$, $(85^\circ,80^\circ)$, $(89^\circ,50^\circ)$, and $(89^\circ,80^\circ)$, respectively. The coloured error bars in each panel represent the uncertainties for different components. The dashed lines represent the equality of model-predicted fractions $f_{\rm true}$ and true fractions $f_{\rm model}$, while the dotted lines are $\pm0.1$ away from the dashed lines.}
    \label{mass-fractions}
\end{figure*}
\begin{figure*}[!htb]
    \centering
    \includegraphics[width=16cm]{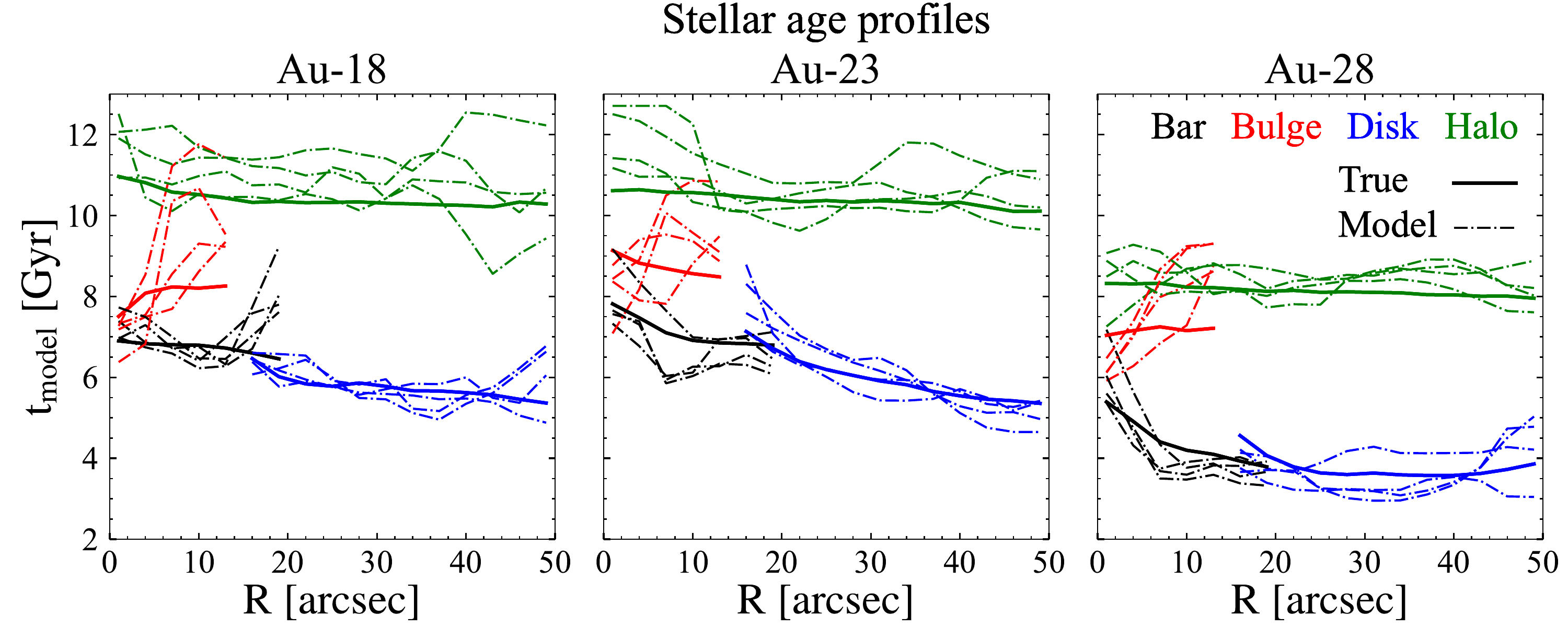}
    \caption{Comparison of the true and model-predicted stellar age profiles of different structures. From left to right, the panels display age profiles for Au-18, Au-23, and Au-28. The black, red, blue, and green lines denote the bar, bulge, disc, and stellar halo components, respectively. The solid lines represent the true mass-weighted age profiles, while the dashed lines denote the corresponding model-predicted profiles derived from mock datasets with different viewing angles.}
    \label{age-profiles}
\end{figure*}
\begin{figure*}[!htb]
    \centering
    \includegraphics[width=16cm]{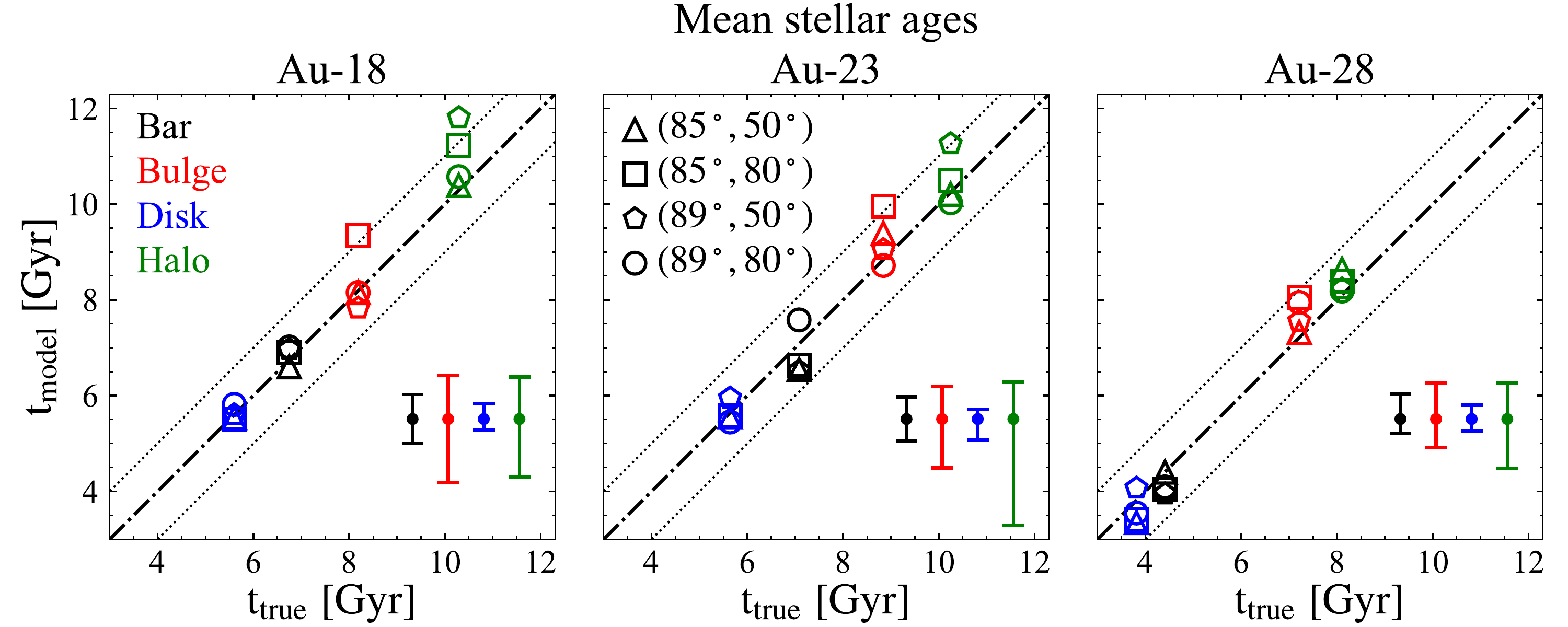}
    \caption{One-to-one comparison of the true and model-predicted mass-weighted mean stellar ages of decomposed structures. The dashed lines represent the equality of model-predicted ages $t_{\rm model}$ and true ages $t_{\rm true}$, while the dotted lines are $\rm\pm1\,Gyr$ away from the dashed lines. The panels and markers are consistent with Fig.~\ref{mass-fractions}.}
    \label{mean-ages}
\end{figure*}
\begin{figure*}
    \centering
    \includegraphics[width=16cm]{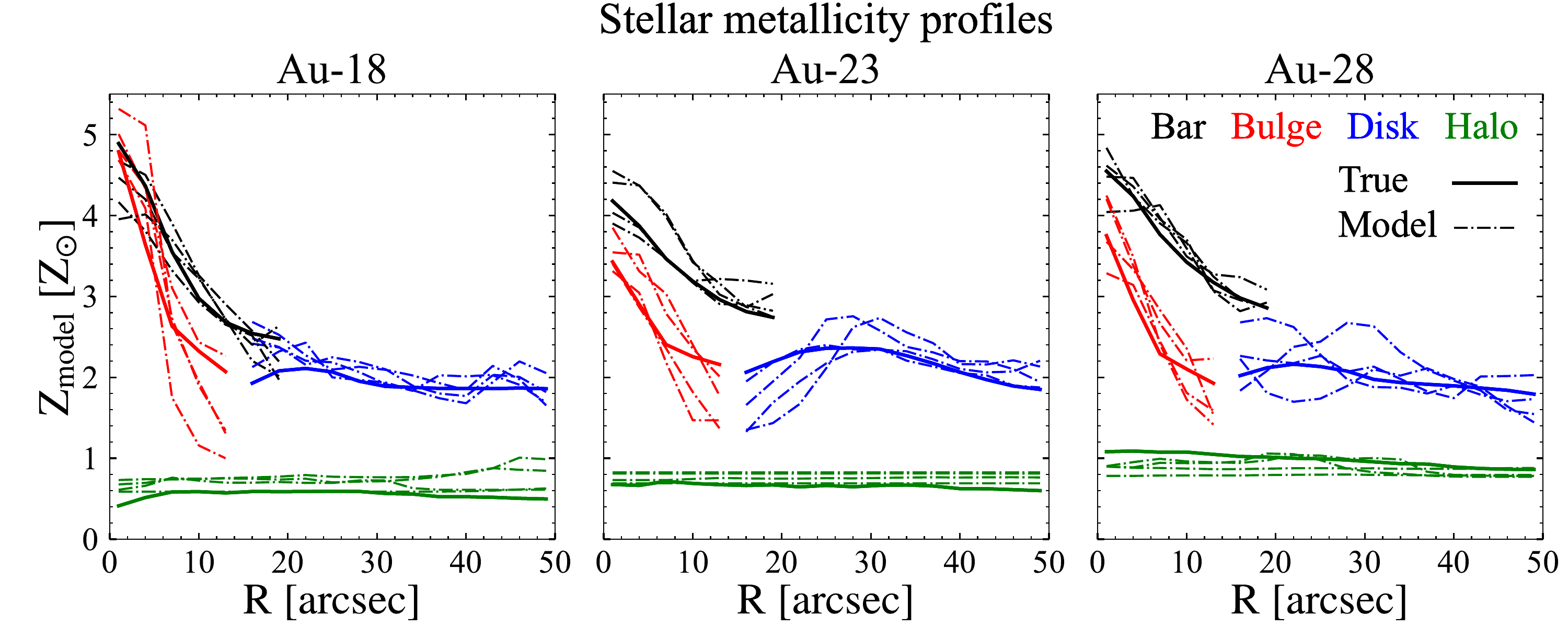}
    \caption{Comparison of the true and model-predicted stellar metallicity profiles of different structures. Each panel follows the same format as Fig.~\ref{age-profiles}.}
    \label{metallicity-profiles}
\end{figure*}
\begin{figure*}
    \centering
    \includegraphics[width=16cm]{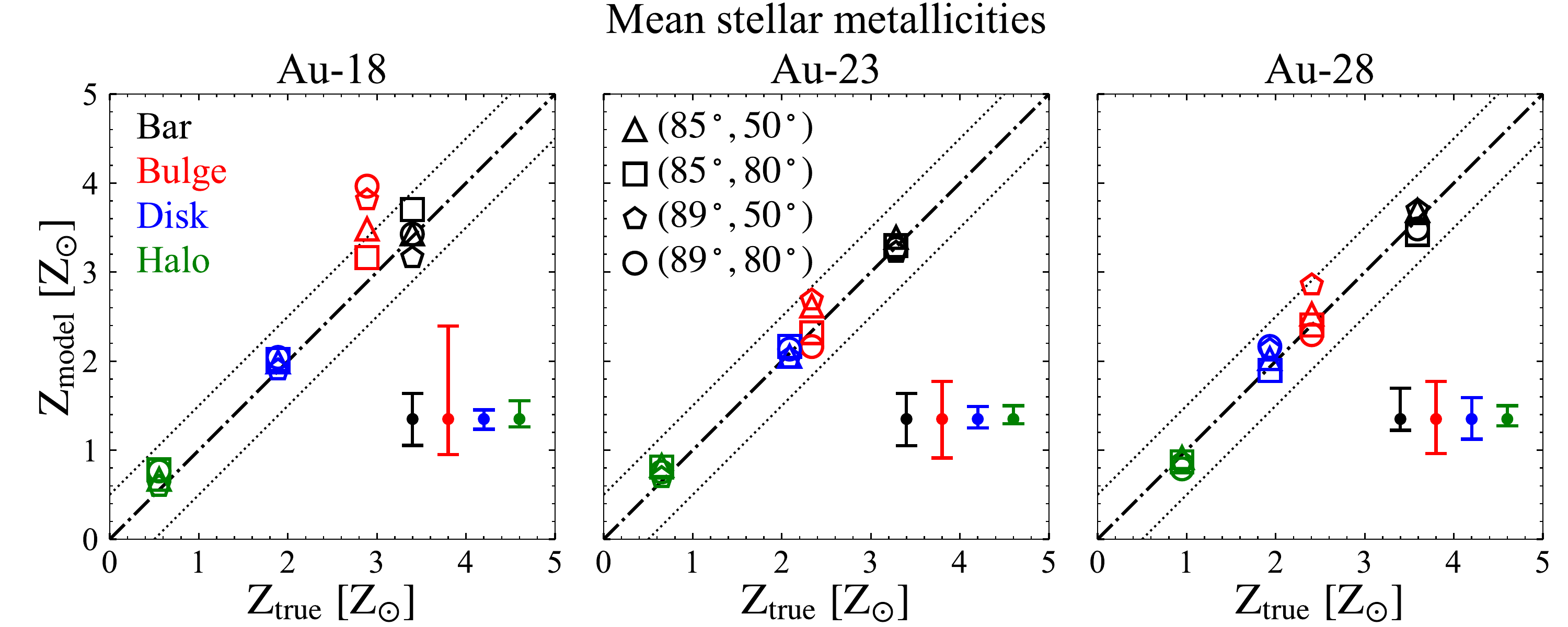}
    \caption{One-to-one comparison of the true and model-predicted mass-weighted mean stellar metallicities of decomposed structures. The dashed lines represent the equality of model-predicted metallicities, $Z_{\rm model}$, and true metallicities, $Z_{\rm true}$, while the dotted lines are $\rm\pm0.5\,Z_{\odot}$ away from the dashed lines. The panels and markers are consistent with Fig.~\ref{mass-fractions}.}
    \label{mean-metallicities}
\end{figure*}

In this section, we use Au-23-85-50 to demonstrate the spatial distributions of stellar mass, age, and metallicity for model-predicted components and compare them with their true counterparts. As mentioned in Sect.~\ref{sec2}, the mock surface brightness was constructed from the stellar mass distribution in simulations with a constant mass-to-light ratio. Therefore, the normalised luminosity and mass distributions are identical.

Using particle information from the simulated galaxy Au-23, we derived the true mass distributions by assuming a constant mass-to-light ratio $M_*/L=2$ (identical to that used when creating the mock surface brightness in Sect.~\ref{sec2}), whereas the model-predicted distributions were calculated from orbit weights and trajectories in the model of Au-23-85-50. By projecting the 3D mass distributions onto the internal $x$-$y$, $x$-$z$, and $y$-$z$ planes, the comparison of true and model-predicted distributions is demonstrated in Fig.~\ref{mass-distributions}. The model successfully reconstructed a BP/X-shaped bar, a spheroidal bulge, a thin disc, and a spatially diffuse stellar halo, closely matching the true structures in all projections. We calculated the mass fraction of each component within the $\rm60\times60\times30\,arcsec^3$ volume. For this model, the mass fractions of all components are recovered with absolute biases $\le0.05$.

Then, we calculated the 3D mass-weighted stellar age distributions for the simulated galaxy Au-23 and the model of Au-23-85-50. We projected the distributions on the $x$-$y$, $x$-$z$, and $y$-$z$ planes, as shown in Fig.~\ref{age-distributions}. The mass-weighted mean stellar ages of the entire galaxy and different components were calculated within the $\rm60\times60\times30\,arcsec^3$ volume. The model reproduced a young disc ($t_{\rm model}=\rm5.5\,Gyr$), an old bulge ($t_{\rm model}=\rm9.4\,Gyr$), an old halo ($t_{\rm model}=\rm10.2\,Gyr$), and an intermediate-age bar ($t_{\rm model}=\rm6.5\,Gyr$). These results generally match the truth, with biases $t_{\rm model}-t_{\rm true}=-0.6$ and $\rm0.6\,Gyr$ for the bar and the bulge, respectively.

Similarly to the age distributions, the mass-weighted stellar metallicity distributions and mean stellar metallicities for Au-23-85-50 are presented in Fig.~\ref{metallicity-distributions}. The model reproduced the most metal-rich bar ($Z_{\rm model}=3.4\,Z_{\odot}$), the most metal-poor halo ($Z_{\rm model}=0.8\,Z_{\odot}$), and the relatively metal-rich bulge and disc ($Z_{\rm model}=2.6$ and $2.1\,Z_{\odot}$, respectively). These values agree with the true metallicities for all components, with all components showing biases, expressed as $|Z_{\rm model}-Z_{\rm true}|\le0.3\,Z_{\odot}$.

\section{Statistical analysis of stellar masses, ages, and metallicities for all cases}
\label{sec5}
\subsection{Uncertainties of model-predicted properties}
\label{sec5.1}
For model-predicted masses from each mock dataset, we estimated their uncertainties following the same approach as in our previous paper \citep{Jin2025}. First, we fixed the orbit libraries and thus also the generated potential in the best-fitting model and randomly perturbed the mock kinematic maps 1000 times based on their error maps. Then, we calculated the orbit weights to derive 1000 new $\chi^2$ values. The fluctuation within the $\pm1\sigma$ ($68\%$; $\Delta\chi^2_{\rm CL}$) region of these $\chi^2$ values is defined as the $1\sigma$ confidence level. The mass variations in models within the $1\sigma$ confidence level ($\chi^2-\chi^2_{\rm min}\le\Delta\chi^2_{\rm CL}$) are treated as uncertainties.

Since the model-predicted stellar ages are derived from the stellar orbits, their uncertainties originate from variations in both mock kinematic maps and age maps. For each mock dataset, we first fixed the age maps and calculated the orbit ages for models within the $1\sigma$ confidence level. The variations among the model-predicted stellar ages are defined as uncertainties induced by the kinematic maps, $\Delta t_{\rm kin}$. Then, we randomly perturbed the mock age maps 1000 times based on their error maps and recalculated the orbit ages with unchanged orbit libraries and weights. The standard deviations of ages from these 1000 perturbations are taken as the uncertainties induced by the age maps, $\Delta t_{\rm age}$. We assumed that these two factors independently contribute to the uncertainties. Thus, the overall uncertainties, $\Delta t_{\rm all}$, could be derived using the error propagation formula ($\Delta t_{\rm all}^2=\Delta t_{\rm kin}^2+\Delta t_{\rm age}^2$). Similarly, for stellar metallicities, the overall uncertainties, $\Delta Z_{\rm all}$, were calculated following the same procedure ($\Delta Z_{\rm all}^2=\Delta Z_{\rm kin}^2+\Delta Z_{\rm met}^2$).

\subsection{Mass density profiles and mass fractions of different structures}
\label{sec5.2}
The mass density profiles derived from all mock datasets are shown in the top panels of Fig.~\ref{mass-density-profiles}. The mass profiles of bars and bulges are compact, while those of discs and halos are extended. There are obvious `holes' in the inner disc, which are distinct from the exponential profiles widely used in photometric decompositions. Across all cases, our models aptly recovered the mass profiles in inner bars, inner bulges, outer discs, and outer halos. However, some uncertainties emerge in outer bars, outer bulges, inner discs, and inner halos, which are due to the low mass contributions at these radii.

To statistically quantify the model results across all mock datasets, we present a one-to-one comparison of the true and model-predicted mass fractions in Fig.~\ref{mass-fractions}. These mass fractions have been calculated within the $\rm60\times60\times30\,arcsec^3$ volume. The halo fractions are recovered with absolute biases of $|f_{\rm model}-f_{\rm true}|\le0.03$ in all cases. For the bulge fractions, 10 out of 12 cases show biases below 0.05, while the other two cases remain below 0.1. The absolute biases for the bar and disc fractions reach maximums of 0.15 and 0.14, respectively. On average, the models overestimate the disc fractions by $f_{\rm model}-f_{\rm true}=0.06$ and underestimate the bar fractions by $f_{\rm model}-f_{\rm true}=-0.06$, while the bulge and halo fractions do not exhibit significant biases. These discrepancies are mainly attributed to the model biases of the dynamical bar lengths, which have been adopted to separate the bar+bulge and disc+halo components.

\subsection{Stellar age profiles and mean stellar ages of different structures}
\label{sec5.3}
Using the weights and ages assigned to the stellar orbits in the modelling, we calculated the mass-weighted stellar age profiles for different structures, as shown in Fig.~\ref{age-profiles}. For bars and discs, the models generally reproduced their negative age gradients. The bars and discs exhibit a consistently negative age gradient, indicating the disc origin of the bar; this trend is generally matched by our model. For bulges and halos, the predicted gradients have larger variations across different mock datasets, indicating higher uncertainties. However, the age difference between the bars and the bulges can still be identified: all three galaxies have bulges older than their bars at all radii, with the difference being the largest for Au-28 because its bar was recently formed ($\sim\rm2.8\,Gyr$; \citealp{Fragkoudi2025}). 

We calculated the mass-weighted mean stellar ages of the decomposed structures within the $\rm60\times60\times30\,arcsec^3$ volume and then compared them with the true values in Fig.~\ref{mean-ages}. Although the models slightly underestimate the bar fractions and overestimate disc fractions, the mean stellar ages of different structures are still statistically aligned with the true values. The mean age biases $t_{\rm model}-t_{\rm true}$ for the bar, bulge, disc, and halo components are $-0.1$, 0.4, $-0.1$, and $\rm0.4\,Gyr$, respectively. We have relatively large uncertainties on the ages of the bulge and halo. One reason is that they are relatively minor components in their galaxies with lower mass fractions. Another reason is that they are older stellar populations and we assigned larger age uncertainties to them when creating the mock maps. For all components, the absolute biases of the stellar ages are $\rm\lesssim1\,Gyr$.

\subsection{Stellar metallicity profiles and mean stellar metallicities of different structures}
\label{sec5.4}
Similarly to Fig.~\ref{age-profiles}, the mass-weighted stellar metallicity profiles for different structures are shown in Fig.~\ref{metallicity-profiles}. For the bars and bulges, our models successfully reproduced their steep negative metallicity gradients and can distinguish their metallicity differences. For the discs and halos, our models generally recovered their flat or weakly positive metallicity gradients within the inner regions ($R\lesssim 30\,\rm arcsec$) and their flat or weakly negative gradients in the outer regions ($R\gtrsim 30\,\rm arcsec$).

 We compare the true and model-predicted mass-weighted mean stellar metallicities within the $\rm60\times60\times30\,arcsec^3$ volume in Fig.~\ref{mean-metallicities}, in a similar way to Fig.~\ref{mean-ages}. The mean metallicity biases $Z_{\rm model}-Z_{\rm true}$ are 0.1, 0.3, 0.1, and $\rm0.1\,Z_{\odot}$, respectively. The models maintain the absolute biases of the stellar metallicities within $\rm\sim0.5\,Z_{\odot}$ for the bar, disc, and halo components. However, the bulge metallicities of Au-18 tend to be overestimated (up to $1.1\,Z_{\odot}$). This discrepancy is mainly caused by the models' underestimation of the bulge mass in outer regions ($R\gtrsim 10\,\rm arcsec$), where the stellar populations are significantly more metal-poor than those in the central bulge regions ($R\lesssim 10\,\rm arcsec$).

\section{Discussions}
\label{sec6}
\subsection{Method improvements and remaining challenges}
\label{sec6.1}
In this study, we combined the barred orbit-superposition method \citep{Tahmasebzadeh2021,Tahmasebzadeh2022} with the population-orbit superposition method \citep{Zhu2020} and extended the combined new method to model edge-on barred galaxies with effective improvements. By incorporating both the morphologies and kinematics of stellar orbits into the structural decomposition framework, our method achieves higher accuracy in recovering different structures ($|f_{\rm model}-f_{\rm true}|\lesssim0.1$), compared to earlier non-edge-on barred galaxy models \citep{Tahmasebzadeh2022}, where the biases for bar and bulge fractions reached $\sim0.15$. Additionally, when tagging the stellar orbits with ages and metallicities, we did not use the age-circularity and age-metallicity correlations adopted in previous works \citep{Zhu2020,Ding2023,Jin2024}. This allowed us to avoid strong assumptions in the modelling. Despite weaker model constraints, our method still provides robust recoveries for stellar populations in different components, which reduces the reliance on prior assumptions and allows flexible studies.

As shown in Fig.~\ref{mass-fractions}, the mass fractions of bar and disc exhibit larger biases and uncertainties than the bulge and halo components. This is attributed to uncertainties of the model-predicted dynamical bar lengths, $R_{\rm bar,model}$. The values of the true and model-predicted bar lengths are shown in Table~\ref{bar-length}. For edge-on observations, it is hard to distinguish between the bar end and the disc. Both the bar end and the disc show similar luminosity distributions from edge-on views and are rotation-dominated ($\lambda_z\sim1$; see Fig.~\ref{orbit-distributions}). The key difference lies in their stellar orbits: disc orbits tend to be round in the $x$-$y$ plane ($p_{\rm orb}\sim1$), while bar orbits elongate along the bar ($p_{\rm orb}\ll1$). However, such a difference in orbit morphologies is indistinguishable with data from edge-on views, which results in the uncertainties we tend to see on dynamical bar lengths.

Previous studies have shown that the presence of a bar produces a high-velocity tail in line-of-sight velocity distributions, corresponding to a positive $h_3$--$V$ correlation (e.g. \citealp{Bureau2005,Molaeinezhad2016,LiZhaoyu2018}), while disc regions exhibit a negative $h_3$--$V$ correlation. Therefore, the transition between positive and negative $h_3$--$V$ correlations could indicate the bar length. For example, the true dynamical bar length of Au-23 ($24.6\,\rm arcsec$; see Table~\ref{bar-length}) is similar to the transition radius estimated from the $h_3$ map of Au-23-85-50 ($R_{\rm bar,h3}\sim22\,\rm arcsec$; see Fig.~\ref{observational-data}).

Furthermore, the molecular gas traced by CO and the ionised gas traced by H$\rm\alpha$ tend to peak at the end of the bar (e.g. in the co-rotation resonance ring) in addition to the central region of the galaxy (e.g. \citealp{Regan1999,Verley2007,DiazGarcia2020,FraserMcKelvie2020,Maeda2020}). By assuming the distribution of emission lines as a ring in the co-rotation resonance close to the end of the bar, gas kinematics can be utilised to estimate the bar lengths (e.g. \citealp{GarciaLorenzo2015,Aguerri2015}). Estimating reliable bar lengths from gas kinematics could improve our stellar dynamical modelling for future applications to real galaxies. In our models, by using the true dynamical bar lengths, $R_{\rm bar,true}$, instead of the model-predicted ones, $R_{\rm bar,model}$, to separate the bar+bulge and disc+halo components, we were able to achieve a closer alignment between the mass profiles of all the components with the true profiles (Fig.~\ref{mass-density-profiles-RbarT}), while the biases and uncertainties of the bar and disc mass fractions were reduced (Fig.~\ref{mass-fractions-RbarT}).

\subsection{The coexistence of bars, classical bulges, and nuclear discs}
\label{sec6.2}
The central structures of barred galaxies can be formed by various physical processes. Studies have shown that during the secular evolution of galaxies, bars can channel gas towards galaxies' centres, gradually forming disc-like bulges (e.g. \citealp{Athanassoula2003,Kormendy2004,Gadotti2011}). Additionally, vertical instabilities in bars will result in BP/X-shaped structures (bar buckling; e.g. \citealp{Raha1991,Kuijken1995,Athanassoula2005,Debattista2006,MartinezValpuesta2006}). Both types are collectively termed `pseudo-bulges', which are dominated by rotation. These contrast with the random-motion-dominated `classical bulges', which are elliptical-like and are likely formed by dissipationless collapse or mergers (e.g. \citealp{Bournaud2007,Fisher2008,Gadotti2009,Hopkins2010}). Due to different formation mechanisms, classical bulges are typically older and more metal-poor than pseudo-bulges. The question of whether barred galaxies can host classical bulges is a decades-long debate.

In our own Galaxy, recent observations have revealed dynamically hot, old, and metal-poor stars in the bulge region. Some of these stars likely originated from the ancient thick disc or halo, while others may belong to a small classical bulge (e.g. \citealp{Ness2013,DiMatteo2015,RojasArriagada2017,Zoccali2017,Lucey2021}). This supports the opinion that classical bulges can coexist with bars in galaxies. For external galaxies, some attempts have been made to model classical bulges in barred galaxies (e.g. \citealp{MendezAbreu2014,Erwin2015}), on the basis, however, of photometric decompositions that tend to exhibit large degeneracies. For the simulated galaxies analysed in this paper, the bulges identified through stellar orbits are dynamically hotter, older, and more metal-poor than the bars, which are consistent with the properties of classical bulges. Our models successfully reproduced these properties, demonstrating the method's ability to study the coexistence of bars and classical bulges in external galaxies together with their chemical properties.

Observations have shown that nuclear discs are common in barred galaxies (e.g. \citealp{Gadotti2020,FraserMcKelvie2025}). Four of the 12 edge-on galaxies observed in the GECKOS survey exhibit strong kinematic evidence for the presence of nuclear discs. In particular, the complex $h_3$--$V$ relation is of interest, as it is anti-correlated, correlated, and anti-correlated again in the nuclear disc, bar, and outer disc regions, respectively (\citealp{FraserMcKelvie2025}; see also \citealp{Chung2004,Bureau2005}). This provides an opportunity for dynamical modelling of galaxies incorporating nuclear discs, constrained by MUSE or MUSE-like data. Using the barred orbit-superposition method developed by \citet{Tahmasebzadeh2021,Tahmasebzadeh2022}, the nuclear disc in a non-edge-on galaxy NGC 4371 was successfully detected \citep{Tahmasebzadeh2024}. Although the simulated galaxies we used in this paper lack nuclear discs, our method can effectively identify such structures if present. Nuclear discs are rotation-dominated ($\lambda_z\sim1$) and their time-averaged orbital radii, $R,$ are smaller than the bar lengths, which means that these structures are identifiable in the $\lambda_z$--$R$ phase space.

Overall, our decomposition method allows us to investigate the possible coexistence of bars, classical bulges, and nuclear discs in real observations. By further analysing the ages and metallicities of these structures, we can better understand their formation processes.

\section{Summary}
\label{sec7}
We selected three barred galaxies from the Auriga simulations and created 12 edge-on mock observations in total (four different projections for each galaxy). The orbit-superposition models for these galaxies were constructed in an earlier work and described in our previous paper \citep{Jin2025}. In the current work, we further decomposed these galaxies into bars, bulges, discs, and halos based on stellar orbits. We then tagged the stellar orbits with ages and metallicities to derive the stellar populations of each structure. We compared the properties of model-predicted components with their true counterparts. Our main findings are listed below.

\begin{enumerate}
\item Our models aptly recovered the morphologies, stellar kinematics, ages, and metallicities of different components, including (BP/X-shaped) bars, spheroidal bulges, thin discs, and spatially diffuse stellar halos.

\item The mass fractions of all components can be constrained with absolute biases of $|f_{\rm model}-f_{\rm true}|\le0.15$. For the disc and bar components, the absolute biases reach maximums of 0.15 and 0.14, respectively. For the halo components, the mass fractions were recovered with $|f_{\rm model}-f_{\rm true}|\le0.03$. For the bulge components, 10 out of 12 cases were characterised by $|f_{\rm model}-f_{\rm true}|\le0.05$. The other two cases exhibit $|f_{\rm model}-f_{\rm true}|\le0.10$. 

\item Our models generally recovered the negative age gradients in the bars and discs of the three simulated galaxies, while the model-predicted age gradients for the bulges and halos come with large uncertainties. Nevertheless, the mean stellar ages of all components have been constrained with absolute biases of $|t_{\rm model}-t_{\rm true}|\rm\lesssim1\,Gyr$.

\item Our models reproduced the steep negative metallicity gradients in the bars and bulges of the three simulated galaxies and the flat, weakly negative, or weakly positive gradients in the discs and halos. Apart from the bulge in Au-18, the mean stellar metallicities of all other components are well constrained with absolute biases of $|Z_{\rm model}-Z_{\rm true}|\rm\le0.5\,Z_{\odot}$.

\end{enumerate}

Our method will be applicable to structural analyses of real galaxies in the future, such as the 36 Milky Way-like edge-on galaxies in the MUSE/VLT programme GECKOS \citep{vdSande2024}. These applications will enable investigations of the possible coexistence of bars, classical bulges, and nuclear discs, as well as their formation histories via stellar population distributions.

\begin{acknowledgements} 
We have used simulations from the Auriga Project public data release \citep{Grand2024} available at \url{https://wwwmpa.mpa-garching.mpg.de/auriga/data}. This work is supported by the National Science Foundation of China under Grant No. 12403017. This work is partly supported by the National Science Foundation of China (Grant No. 11821303 to SM) and CAS Project for Young Scientists in Basic Research, Grant No. YSBR-062 (LZ).
\end{acknowledgements}

\begin{appendix}
\onecolumn
\section{Dynamical bar lengths and their influence on model results}
We present the values of the true and model-predicted bar lengths in Table~\ref{bar-length}. As described in Sect.~\ref{sec3.2}, these dynamical bar lengths are defined as the smallest radius where the cold orbit fraction, $f_{\rm cold}\ge0.5$. We also present the bar lengths directly estimated from observed kinematic maps by eyes, which correspond to the transition radii between positive and negative $h_3$--$V$ correlations.
\begin{table*}[!htb]
\centering
\caption{Dynamical bar lengths: true values, model predictions, and estimations from observed maps}
\begin{tabular}{|c|c|c|c|c|}
\hline
Galaxy & $R_{\rm bar,true}$ (arcsec) & $(\theta_{\rm T},\varphi_{\rm T})$ & $R_{\rm bar,model}$ (arcsec) & $R_{\rm bar,h3}$ (arcsec) \\
\hline
\multirow{4}*{Au-18} & \multirow{4}*{20.5} & $(85^\circ,50^\circ)$ & 16.3 & $\sim17$ \\
\cline{3-5}
                   ~ &                   ~ & $(85^\circ,80^\circ)$ & 15.6 & $\sim18$ \\
\cline{3-5}
                   ~ &                   ~ & $(89^\circ,50^\circ)$ & 16.6 & $\sim16$ \\
\cline{3-5}
                   ~ &                   ~ & $(89^\circ,80^\circ)$ & 15.9 & $\sim18$ \\
\hline
\multirow{4}*{Au-23} & \multirow{4}*{24.6} & $(85^\circ,50^\circ)$ & 23.1 & $\sim22$ \\
\cline{3-5}
                   ~ &                   ~ & $(85^\circ,80^\circ)$ & 26.1 & $\sim26$ \\
\cline{3-5}
                   ~ &                   ~ & $(89^\circ,50^\circ)$ & 22.1 & $\sim23$ \\
\cline{3-5}
                   ~ &                   ~ & $(89^\circ,80^\circ)$ & 29.2 & $\sim26$ \\
\hline
\multirow{4}*{Au-28} & \multirow{4}*{22.0} & $(85^\circ,50^\circ)$ & 21.0 & $\sim17$ \\
\cline{3-5}
                   ~ &                   ~ & $(85^\circ,80^\circ)$ & 27.1 & $\sim22$ \\
\cline{3-5}
                   ~ &                   ~ & $(89^\circ,50^\circ)$ & 17.1 & $\sim17$ \\
\cline{3-5}
                   ~ &                   ~ & $(89^\circ,80^\circ)$ & 25.8 & $\sim21$ \\
\hline
\end{tabular}
\tablefoot{From left to right: (1) galaxy name; (2) true dynamical bar length, $R_{\rm bar,true}$; (3) inclination angle, $\theta_{\rm T}$, and bar azimuthal angle, $\varphi_{\rm T}$; (4) model-predicted dynamical bar length, $R_{\rm bar,model}$; (5) bar length, $R_{\rm bar,h3}$, estimated from observed kinematic maps by eye. We note that $\rm 1\,arcsec=0.2\,kpc$.}
\label{bar-length}
\end{table*}

To detect the influence of dynamical bar lengths on model results, we use the true dynamical bar lengths, $R_{\rm bar,true}$, instead of the model-predicted ones, $R_{\rm bar,model}$, to separate the bar+bulge and disc+halo components in the modelling, while keeping other parameters in the decomposition framework unchanged. We present the new model results in Figs.~\ref{mass-density-profiles-RbarT} to~\ref{mean-metallicities-RbarT}. Compared to the original models, the mass profiles of all components align more closely with the true profiles. The biases and uncertainties of the mass fractions are reduced, particularly for the bars and discs. However, the new models show no significant improvement in reproducing the stellar age and metallicity profiles, primarily because the inner disc and outer bar regions ($R\approx 20\,\rm arcsec$) exhibit similar stellar ages and metallicities.

\begin{figure*}[!htb]
    \centering
    \includegraphics[width=16cm]{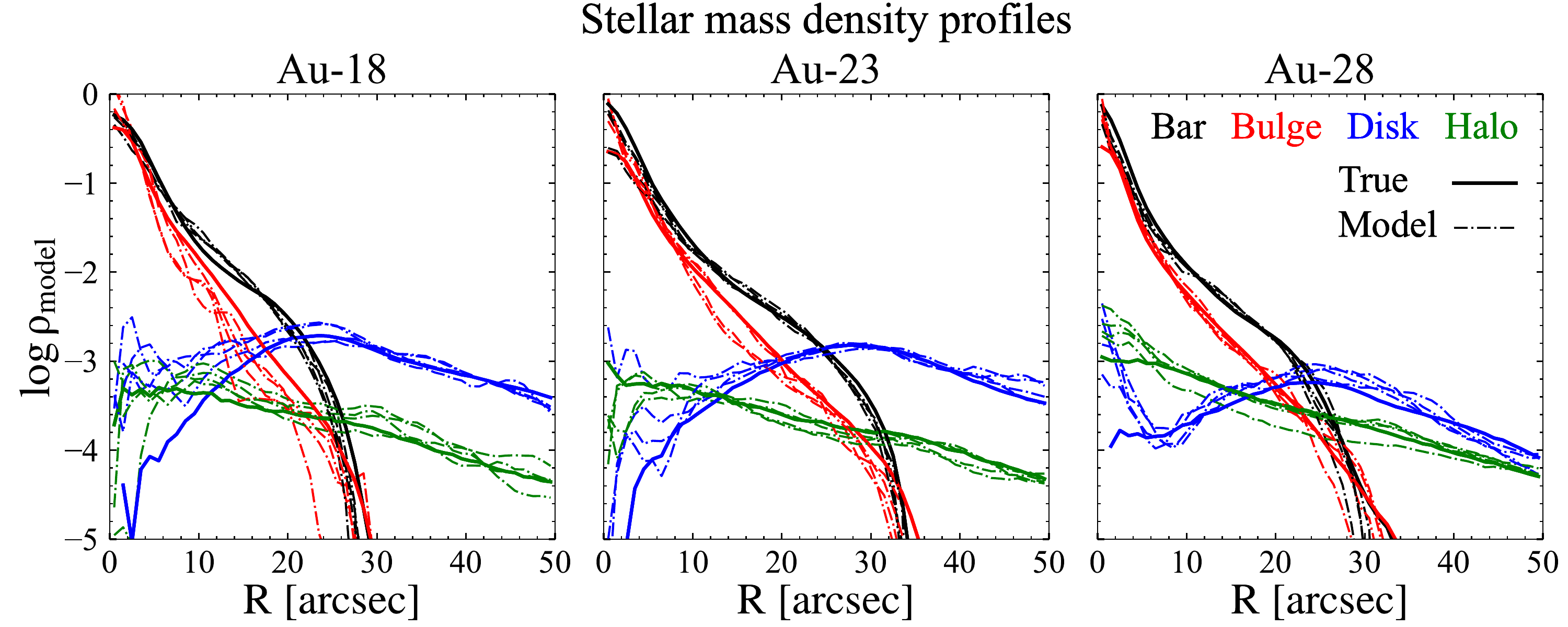}
    \caption{Similarly to Fig.~\ref{mass-density-profiles}, but for model results with true dynamical bar lengths, $R_{\rm bar,true}$.}
    \label{mass-density-profiles-RbarT}
\end{figure*}
\begin{figure*}
    \centering
    \includegraphics[width=16cm]{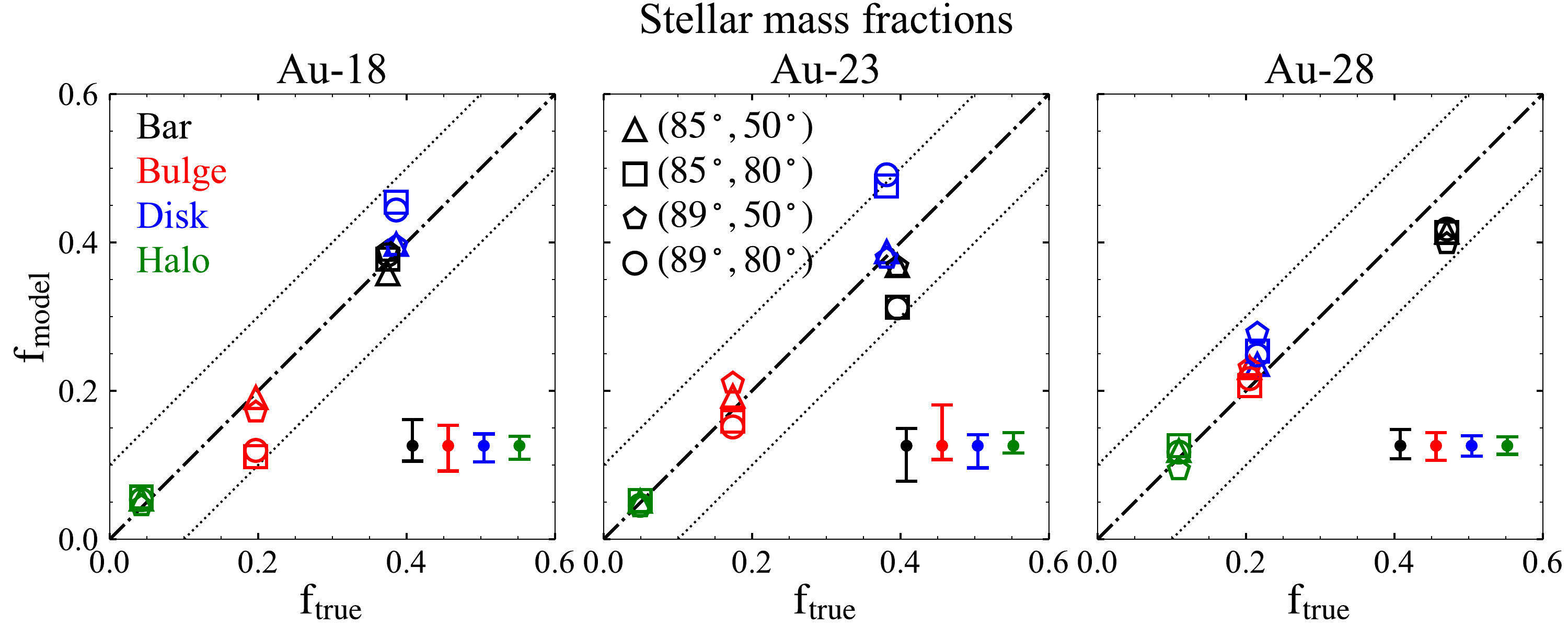}
    \caption{Similarly to Fig.~\ref{mass-fractions}, but for model results with true dynamical bar lengths, $R_{\rm bar,true}$.}
    \label{mass-fractions-RbarT}
\end{figure*}

\begin{figure*}
    \centering
    \includegraphics[width=16cm]{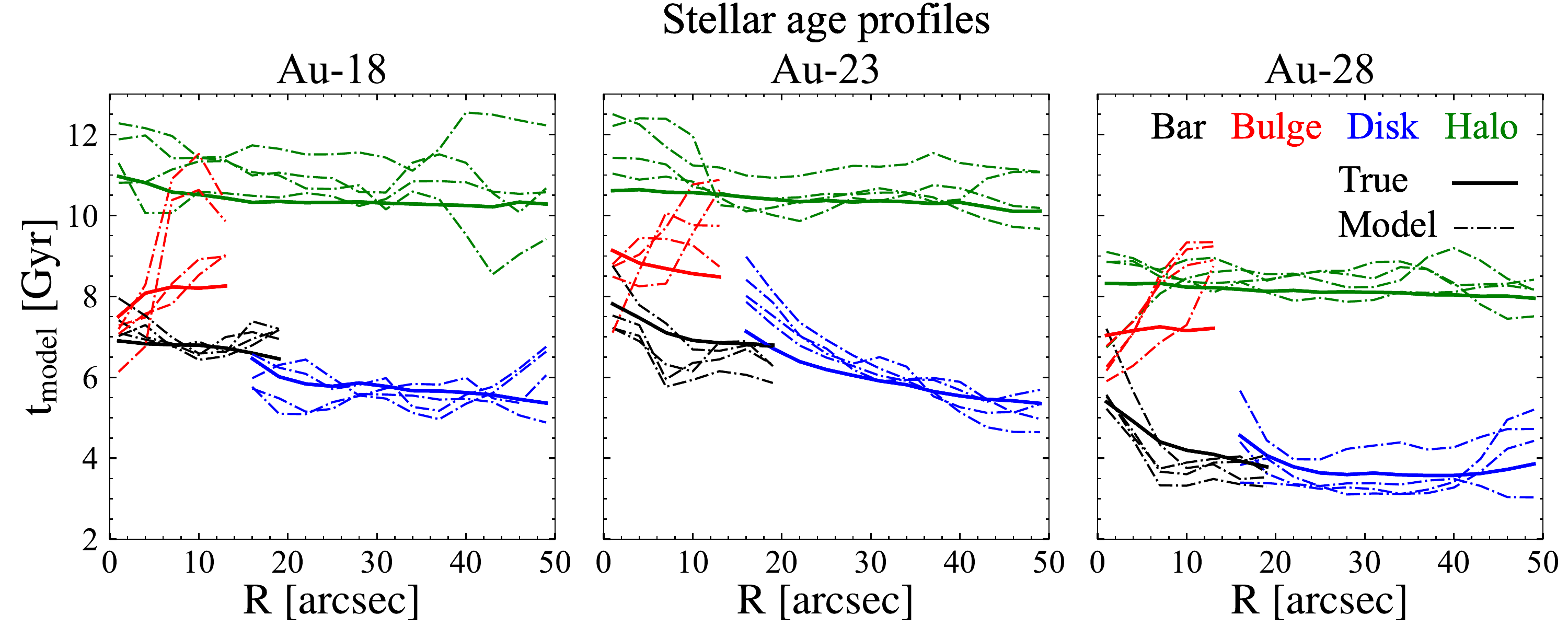}
    \caption{Similarly to Fig.~\ref{age-profiles}, but for model results with true dynamical bar lengths, $R_{\rm bar,true}$.}
    \label{age-profiles-RbarT}
\end{figure*}
\begin{figure*}
    \centering
    \includegraphics[width=16cm]{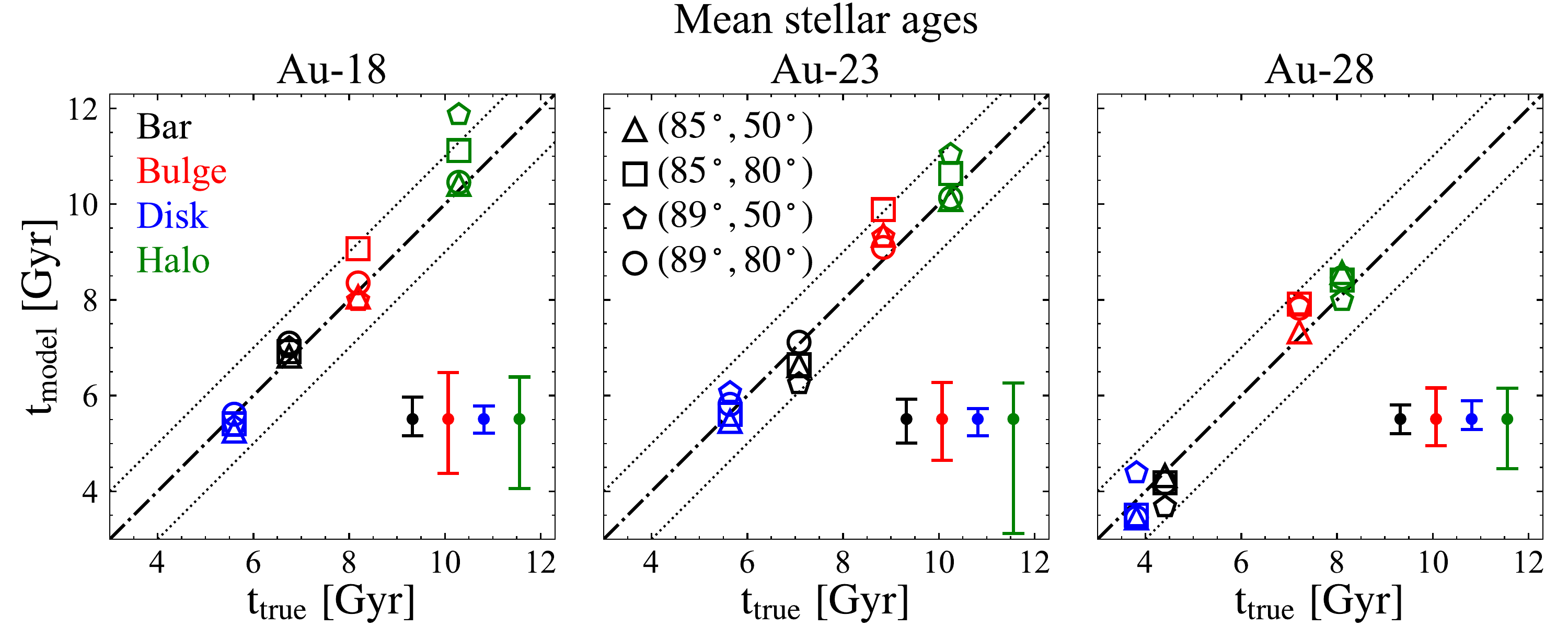}
    \caption{Similarly to Fig.~\ref{mean-ages}, but for model results with true dynamical bar lengths, $R_{\rm bar,true}$.}
    \label{mean-ages-RbarT}
\end{figure*}

\begin{figure*}
    \centering
    \includegraphics[width=16cm]{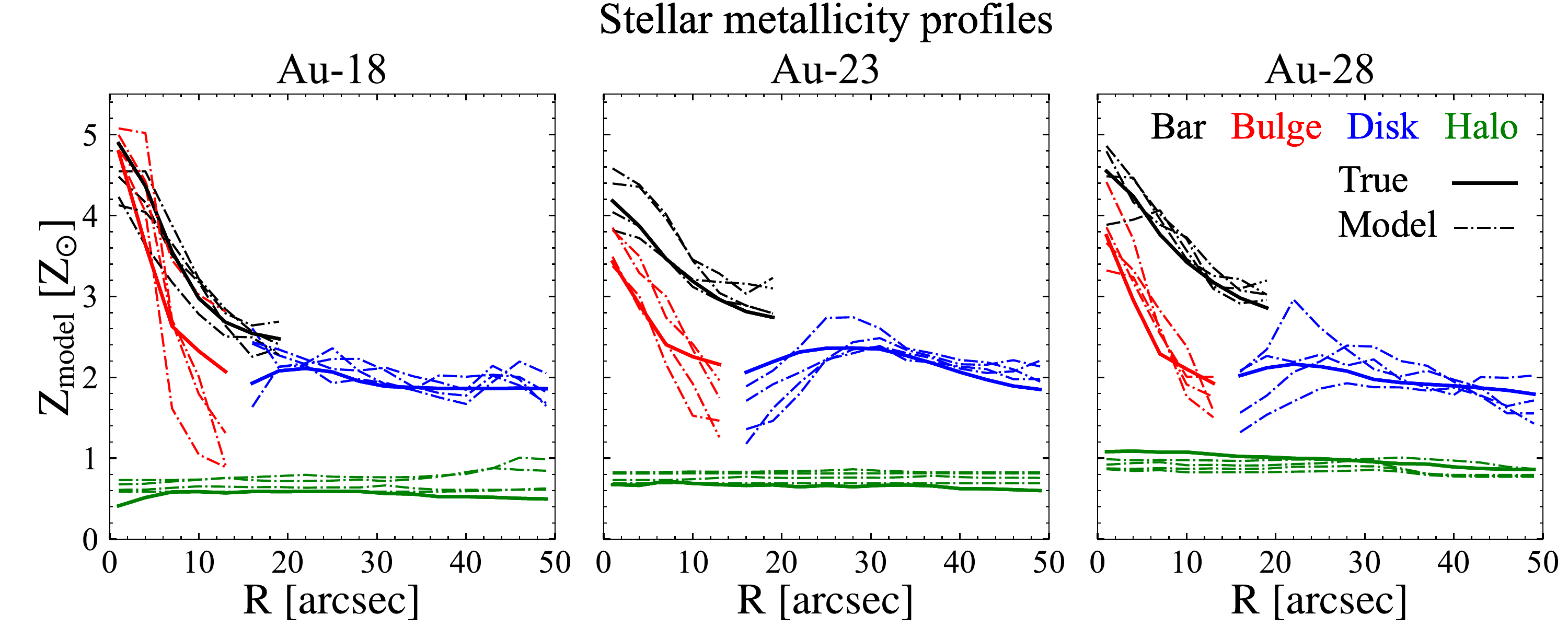}
    \caption{Similarly to Fig.~\ref{metallicity-profiles}, but for model results with true dynamical bar lengths, $R_{\rm bar,true}$.}
    \label{metallicity-profiles-RbarT}
\end{figure*}
\begin{figure*}
    \centering
    \includegraphics[width=16cm]{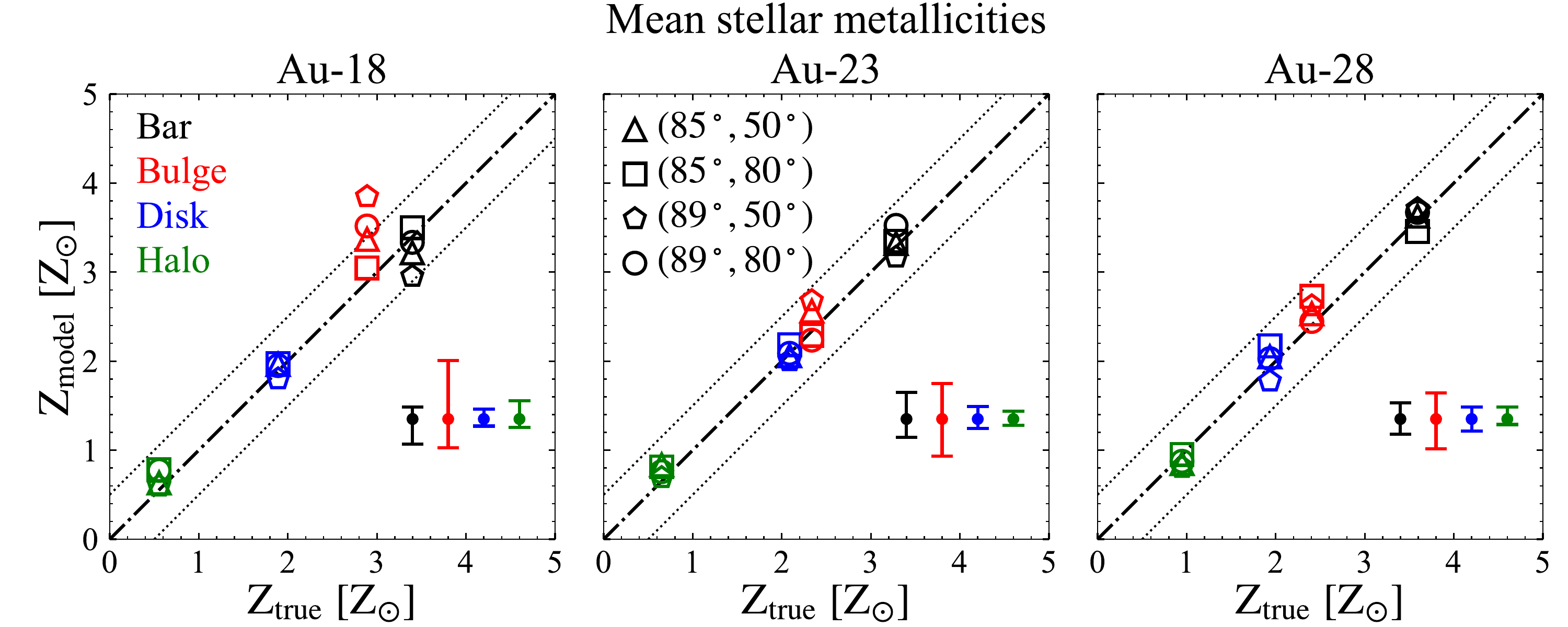}
    \caption{Similarly to Fig.~\ref{mean-metallicities}, but for model results with true dynamical bar lengths, $R_{\rm bar,true}$.}
    \label{mean-metallicities-RbarT}
\end{figure*}

\end{appendix}
\end{nolinenumbers}
\end{document}